\documentclass[12pt]{iopart}
\usepackage{iopams}
\usepackage{amssymb}
\usepackage{graphicx,subfigure,epstopdf}

\usepackage{cite}
\eqnobysec
\usepackage[colorlinks,linktocpage,linkcolor=blue]{hyperref}

\begin{document}

\title[]{Resonant behavior of the generalized Langevin system with tempered Mittag-Leffler memory kernel}

\author{Yao Chen, Xudong Wang, and Weihua Deng}

\address{School of Mathematics and Statistics, Gansu Key Laboratory of Applied Mathematics and Complex Systems, Lanzhou University, Lanzhou 730000, P.R. China}
\ead{ychen2015@lzu.edu.cn, xdwang14@lzu.edu.cn, and dengwh@lzu.edu.cn}
\vspace{10pt}

\begin{abstract}
The generalized Langevin equation describes anomalous dynamics. Noise is not only the origin of uncertainty but also plays a positive role in helping to detect signal with information, termed stochastic resonance (SR). This paper analyzes the anomalous resonant behaviors of the generalized Langevin system with a multiplicative dichotomous noise and an internal tempered Mittag-Leffler noise.  For the system with fluctuating harmonic potential, we obtain the exact expressions of several SR, such as, the first moment, the amplitude and the autocorrelation function for the output signal as well as the signal-noise ratio. We analyze the influence of the tempering parameter and memory exponent on the bona fide SR and the general SR. Moreover, it is detected that the critical memory exponent changes regularly with the increase of tempering parameter. Almost all the theoretical results are validated by numerical simulations. \\

%

\noindent Keywords: generalized Langevin equation, tempered Mittag-Leffler friction memory kernel, dichotomous noise, stochastic resonance, amplitude, signal-noise ratio\\

\noindent (Some figures may appear in colour only in the online journal)
\end{abstract}

\section{Introduction}
The fundamental equation for approximating the dynamics of nonequilibrium system is (generalized) Langevin equation, which contains both frictional and random forces, being related by fluctuation-dissipation theorem \cite{Zwanzig:2001}. Extending the frictional term of standard Langevin equation to have memory kernel leads to the generalized Langevin equation (GLE),
\begin{equation}\label{har_osci}
\ddot{x}(t)+\int_0^t\eta(t-t')\dot{x}(t')\rmd t'=\xi(t),
\end{equation}
where $x(t)$ is the displacement and $\eta(t)$ is the dissipation memory kernel linking to the autocorrelation function of the internal noise $\xi(t)$ through the second fluctuation-dissipation theorem \cite{Kubo:1966}: $\langle\xi(t_1)\xi(t_2)\rangle=k_BT\eta(t_1-t_2)$ with the Boltzmann constant $k_B$ and the absolute temperature $T$.

The GLE with power law memory kernel describes anomalous dynamics \cite{Deng:2009,Lutz:2001,Burov:2008}, mainly focusing on subdiffusion. A Mittag-Leffler correlated random force is introduced to GLE in \cite{Vinales:2007}, which can describe both subdiffusion and superdiffusion and the autocorrelation function of which is nonsingular at origin and behaves as a power law for large $t$. More specifically, the Mittag-Leffler memory kernel is defined as
\begin{equation}\label{ML}
\eta(t)=\frac{\gamma}{\tau^{\alpha}}E_{\alpha}\left(-\frac{t^{\alpha}}{\tau^{\alpha}}\right),~~~t\geq0,
\end{equation}
where $E_{\alpha}(t)=\sum_{k=0}^{\infty}\frac{t^k}{\Gamma(\alpha k+1)}$ denotes the Mittag-Leffler function \cite{Erdelyi:1981}. In (\ref{ML}), $0<\alpha<2$ is the memory exponent, $\tau$ is the characteristic memory time, and $\gamma$ is the friction coefficient. 
Taking $\alpha=1$, (\ref{ML}) becomes an exponential form
\begin{equation}\label{exp}
\eta(t)=\frac{\gamma}{\tau}\textrm{exp}\left(-\frac{t}{\tau}\right).
\end{equation}
In this case, (\ref{har_osci}) describes the process driven by the Ornstein-Uhlenbrck noise \cite{Wang:1996}. Moreover, in the limit $\tau\rightarrow0$, the colored noise (\ref{exp}) reduces to Gaussian white noise $\eta(t)=2\gamma\delta(t)$ \cite{Vinales:2009}. On the other hand, for $\alpha\neq1$, we can write (\ref{ML}) as the following asymptotic forms \cite{Lutz:2001,Pottier:2011}:
\begin{equation}\label{smallt}
\eta(t)\simeq\frac{\gamma}{\tau^{\alpha}}\textrm{exp}\left(-\frac{(t/\tau)^{\alpha}}{\Gamma(\alpha+1)}\right),~~~t\ll\tau,
\end{equation}
and
\begin{equation}\label{larget}
\eta(t)\simeq\frac{\gamma}{\tau^{\alpha}}\frac{1}{\Gamma(1-\alpha)}\left(\frac{t}{\tau}\right)^{-\alpha},~~~t\gg\tau,
\end{equation}
which means that the correlation behaves as a power law in the asymptotic limit at a short memory time $\tau$. In the case of (\ref{larget}), the system (\ref{har_osci}) displays subdiffusion for $0<\alpha<1$ and superdiffusion for $1<\alpha<2$ \cite{Vinales:2007}.


Noise is an extremely important concept; it means a lot of things in science and engineering; an  issue that naturally comes to our mind is to extract desired information (signal) from a background of unwanted noise in an ultrasound machine. Sometimes, noise is no trouble at all; on the contrary, it can help to amplify weak useful signal, termed stochastic resonance (SR). SR was originally coined by Benzi and collaborators to explain the periodic recurrence of the ice ages \cite{Benzi:1981}. The review paper \cite{Gammaitoni:1998} points out that SR has three basic ingredients: a form of threshold; a weak coherent input; and a source of noise inherent
in the system or being added to the coherent input. The original understanding of SR is that it can occur only in nonlinear system, but in recent years, the SR has been found in linear system with multiplicative noise \cite{Gitterman:1999,Gitterman:2004}. It is often observed that when a physical system is displaced from its equilibrium position, the system experiences a restoring force; if the force is proportional to the displacement, the system is called a harmonic oscillator, being widely studied in physics \cite{Lindenberg:1980,Landa:1996}. The anomalous diffusive behavior of (\ref{har_osci}) with harmonic oscillator is discussed in \cite{Burov:20080} for power law noise and in \cite{Vinales:2009} for Mittag-Leffler noise. In particular, in \cite{Vinales:2009}, it is found that the system with Mittag-Leffler noise exhibits more oscillations than the one with pure power law noise.

Adding a periodic input signal (external periodic driving force) $A_0\textrm{cos}(\Omega t)$ and the fluctuating harmonic potential $V(x,t)=(\omega^2+z(t))\frac{x^2}{2}$ to (\ref{har_osci}) leads to the type of system we consider in this paper \cite{Soika:2012,Soika:2010}:
\begin{equation}\label{SR}
\ddot{x}(t)+\int_0^t\eta(t-t')\dot{x}(t')\rmd t'+\omega^2x(t)+z(t)x(t)=A_0\textrm{cos}(\Omega t)+\xi(t),
\end{equation}
where $A_0$ is the amplitude, $\Omega$ is the driving angular frequency, $\omega$ is undamped angular frequency, and $z(t)$ is a multiplicative noise \cite{Gammaitoni:1994,Schenzle:1979,Jung:1984}, playing the role of fluctuating control parameter. In fact, the multiplicative fluctuations are usual in nature, especially in biological systems \cite{Astumian:1996}. The model (\ref{SR}) involving power law or Mittag-Leffler memory kernel is discussed in \cite{Soika:2012,Soika:2010,Laas:2011,He:2017} with the concern on how the behavior of response function (output amplitude) depends on the system parameters. Under some circumstances, for more accurately/reasonabely approximating physical problems, e.g., considering the particles' finite lifespan and the restricted moving space, tempered anomalous dynamics are widely considered \cite{Wu:2016,Sandev:2015,Stanislavsky:2008,Meerschaert:2013,Chen:2017,Liemert:2017}. This paper discusses the SR phenomenon of the model (\ref{SR}) with tempered Mittag-Leffler memory kernel defined in (\ref{TML}), and uncovers how the behavior of the output amplitude depends on the system parameters and the influence of exponential truncation on the resonance behavior. We also analyze signal-noise ratio (SNR), its dependence on the system parameters, and the influence of tempering parameter. Obtaining the theoretical results requires complex analysis and sophisticated mathematical tools; to verify their correctness, the numerical simulations are performed by Monte Carlo methods with algorithms presented in Appendix.

The outline of this paper is as follows. In Section \ref{two}, we present the model which describes the oscillator system with tempered Mittag-Leffler frictional memory kernel, and obtain the first moment of the output signal as well as its amplitude. In Section \ref{three}, the dependence of the output signal's amplitude on the system parameters, such as the characteristic memory time $\tau$, the driving frequency $\Omega$, and the amplitude $a$ of the multiplicative noise, is presented;  besides, we give the conditions that the friction coefficient $\gamma$ and the memory exponent $\alpha$ ($0<\alpha<2$) should satisfy for the appearance of resonance phenomenon; moreover, the exponential tempering parameter affects the critical memory exponent and reduces the resonance region in parameter space $(\gamma, \alpha)$. In Section \ref{four}, we calculate the correlation function of the output signal and the SNR of the oscillator dynamic system, and consider SNR versus driving frequency $\Omega$ for different truncation parameters and memory exponents. Finally, in Section \ref{five}, some conclusions are presented.

\section{Model}\label{two}
In the dynamic oscillatory system (\ref{SR}), which is susceptible to the noisy environment,
we take into account the tempered Mittag-Leffler friction memory kernel,  i.e., a modification to Mittag-Leffler memory kernel by an exponential truncation:
\begin{equation}\label{TML}
\eta(t)=\frac{\gamma}{\tau^{\alpha}}\textrm{e}^{-bt}E_{\alpha}\left(-\frac{t^{\alpha}}{\tau^{\alpha}}\right),~t\geq0,~0<\alpha<2,
~b\geq0,~\gamma>0,~\tau>0.
\end{equation}
Here $b$ is the truncation parameter, and the other parameters are the same as the ones in (\ref{ML}). We consider a dichotomically perturbed oscillator added on the harmonic potential $V(x)=\omega^2\frac{x^2}{2}$, which is determined by the dichotomous noise $z(t)$ \cite{Horsthemke:1984} consisting of jumps between two values: $z_1=a$ and $z_2=-a$ with $a>0$, and the value of the jump occurs with the stationary probabilities $P_s(a)=P_s(-a)=\frac{1}{2}$ and follows the pattern of a Poisson process in time. In fact, the dichotomous noise $z(t)$ can be completely characterized by the following transition probability because of its time-homogeneous Markovian property:
\begin{equation*}
  P_{ij}(t)=\frac{1}{2}\left(\begin{array}{cc} 1+\textrm{e}^{-\nu t}  &  1-\textrm{e}^{-\nu t} \\ 1-\textrm{e}^{-\nu t}  & 1+\textrm{e}^{-\nu t} \end{array} \right), ~~~i, j\in\{-a, a\},
\end{equation*}
with the switching rate $\nu>0$. 
Following the transition probability, it can be calculated that the mean value of dichotomous noise $\langle z(t)\rangle=0$ and the autocorrelation function $\langle z(t+\tau)z(t)\rangle=a^2\textrm{e}^{-\nu |\tau|}$. Moreover, we assume that $z(t)$ and $\xi(t)$ are independent with $\langle z(t)\xi(t)\rangle=0$.

To study the SR phenomenon, we mainly discuss the behavior of the first moment $\langle x(t)\rangle$ of the system (\ref{SR}). The usual approach of getting the exact expression of $\langle x(t)\rangle$ is to use the Shapiro-Loginov formula \cite{Shapiro:1978} :
\begin{equation}
\frac{\rmd}{\rmd t}\langle z(t)g(z)\rangle=\left\langle z(t)\frac{\rmd}{\rmd t}g(z)\right\rangle-\nu\langle  z(t)g(z)\rangle,
\end{equation}
where $g(z)$ is an arbitrary functional of noise $z(t)$. More precisely, define $X_1=\langle x(t)\rangle$, $X_2=\langle \dot{x}(t)\rangle$, $X_3=\langle z(t)x(t)\rangle$, and $X_4=\langle z(t)\dot{x}(t)\rangle$. Then from (\ref{SR}) and (\ref{TML}), there exists
\begin{eqnarray}\label{x1}
    \dot{X}_1=X_2, \nonumber\\
    \dot{X}_2=-\omega^2X_1-X_3-\int_0^t\eta(t-t')X_2(t')\rmd t'+A_0 \textrm{cos}(\Omega t),  \nonumber\\
    \dot{X}_3=X_4-\nu X_3, \nonumber\\
    \dot{X}_4=-a^2X_1-\omega^2X_3-\nu X_4- \textrm{e}^{-\nu t}\int_0^t\eta(t-t')X_4(t')\textrm{e}^{\nu t'}\rmd t',
\end{eqnarray}
where the facts that $\langle z^2(t)x(t)\rangle=a^2\langle x(t)\rangle$ and $\langle \dot{x}(t')z(t)\rangle=\langle \dot{x}(t')z(t')\rangle \textrm{e}^{-\nu (t-t')}$ have been used.
Using the Laplace transform technique, we obtain
\begin{eqnarray*}
    sX_1(s)=X_2(s)+X_1(0), \\
    (s+\eta(s))X_2(s)=-\omega^2X_1(s)-X_3(s)+A_0\mathcal{L}[\textrm{cos}(\Omega t)]+X_2(0), \\
    (s+\nu)X_3(s)=X_4(s)+X_3(0), \\
    (s+\nu+\eta(s+\nu))X_4(s)=-a^2X_1(s)-\omega^2X_3(s)+X_4(0),
\end{eqnarray*}
where $X_i(s)$ is the Laplace transform of $X_i(t)$, $i=1,2,3,4$, i.e., $X_i(s)=\int_0^\infty \textrm{e}^{-st}X_i(t)\rmd t$. 
Solving the above equations leads to
\begin{equation}\label{solution1}
X_1=\sum_{k=1}^4 H_{1k}(t)X_k(0)+A_0\int_0^tH_{12}(t-t')\textrm{cos}(\Omega t')\rmd t',
\end{equation}
and
\begin{equation}\label{solution2}
X_3=\sum_{k=1}^4 H_{3k}(t)X_k(0)+A_0\int_0^tH_{32}(t-t')\textrm{cos}(\Omega t')\rmd t'
\end{equation}
with
\begin{eqnarray}
&H_{11}(s)=[\omega^2+(s+\nu)^2+(s+\nu)\eta(s+\nu)]\frac{s+\eta(s)}{D(s)}, \nonumber\\
&H_{12}(s)=\frac{\omega^2+(s+\nu)^2+(s+\nu)\eta(s+\nu)}{D(s)}, \nonumber\\
&H_{13}(s)=-\frac{s+\nu+\eta(s+\nu)}{D(s)}, \nonumber\\
&H_{14}(s)=-\frac{1}{D(s)}, \nonumber\\
&H_{31}(s)=-\frac{a^2[s+\eta(s)]}{D(s)}, \nonumber\\
&H_{32}(s)=-\frac{a^2}{D(s)}, \nonumber\\
&H_{33}(s)=\frac{[\omega^2+s^2+s\eta(s)][s+\nu+\eta(s+\nu)]}{D(s)}, \nonumber\\
&H_{34}(s)=\frac{\omega^2+s^2+s\eta(s)}{D(s)}, \nonumber\\[3pt]
&D(s)=[\omega^2+s^2+s\eta(s)][\omega^2+(s+\nu)^2+(s+\nu)\eta(s+\nu)]-a^2, \nonumber\\[3pt]
&\eta(s)=\frac{\gamma}{\tau^{\alpha}}\frac{(s+b)^{\alpha-1}}{(s+b)^\alpha+\tau^{-\alpha}}.\nonumber
\end{eqnarray}


In order to ensure the stability of the solutions (\ref{solution1}) and (\ref{solution2}),  the roots of $D(s)=0$ should not have positive real part. To meet this condition,
\begin{equation}\label{stability}
a^2<a^2_{cr}=\omega^2\left[\omega^2+\nu^2+\frac{\gamma \nu (\nu+b)^{\alpha-1}}{\tau^\alpha(\nu+b)^\alpha+1}\right]
\end{equation}
or $a^2=a^2_{cr}$. Assuming the inequality (\ref{stability}) is fulfilled, in the long time limit, the influence of the initial conditions can be ignored, i.e.,
\begin{equation}\label{fir_mon}
\langle x(t)\rangle_{as}=A_0\int_0^tH_{12}(t-t')\textrm{cos}(\Omega t')\rmd t'
\end{equation}
and
\begin{equation}\label{cor}
\langle z(t)x(t)\rangle_{as}=A_0\int_0^tH_{32}(t-t')\textrm{cos}(\Omega t')\rmd t'.
\end{equation}
Finally, using the techniques in \cite{Gammaitoni:1998,Hanggi:1978,Hanggi:1982,Hanggi:1977}, the asymptotic expressions of the first moments (\ref{fir_mon}) and (\ref{cor}) read
\begin{equation}\label{fir_mon_as}
\langle x(t)\rangle_{as}=\textrm{sgn}(\chi'(\Omega))A\, \textrm{cos}(\Omega t+\Psi),
\end{equation}
\begin{equation}\label{cor_as}
\langle z(t)x(t)\rangle_{as}=\textrm{sgn}(\chi'_1(\Omega))A_1 \textrm{cos}(\Omega t+\Psi_1),
\end{equation}
where the variables above will be explicitly explained in the following. In (\ref{fir_mon_as}),
 the output amplitude $A=A_0|\chi(\Omega)|$, the phase shift $\Psi= \textrm{arctan}(-\frac{\chi''}{\chi'})$, the sign function depends on the choice of $\Psi$, and the complex susceptibility $\chi(\Omega)$ of the dynamical oscillator system is
\begin{equation*}
\chi(\Omega)=\chi'(\Omega)+\rmi \chi''(\Omega)=H_{12}(-\rmi \Omega)=\int_0^\infty \textrm{e}^{\rmi\Omega t}H_{12}(t)\rmd t.
\end{equation*}
Here
\begin{eqnarray*}
\chi(\Omega)=&\frac{m_1(m^2+n^2)-a^2m}{(m_1m-n_1n-a^2)^2+(m_1n+n_1m)^2}\\
&+\rmi\frac{n_1(m^2+n^2)+a^2n}{(m_1m-n_1n-a^2)^2+(m_1n+n_1m)^2},
\end{eqnarray*}
\begin{equation}\label{A}
A=A_0\frac{(m^2+n^2)^{\frac{1}{2}}}{((m_1m-n_1n-a^2)^2+(m_1n+n_1m)^2)^{\frac{1}{2}}},
\end{equation}
and
\begin{equation*}
\Psi=\textrm{arctan}\left(\frac{a^2n+n_1(m^2+n^2)}{a^2m-m_1(m^2+n^2)}\right),
\end{equation*}
with
\begin{eqnarray*}
\varphi=\textrm{arctan}\left(\frac{\Omega}{\nu+b}\right), \quad
\theta=\textrm{arctan}\left(\frac{\Omega}{b}\right), \quad
\phi=\textrm{arctan}\left(\frac{\Omega}{\nu}\right), \\[3pt]
f=\frac{(\Omega^2+(\nu+b)^2)^{\frac{\alpha}{2}} \cos(\phi-\varphi)\tau^\alpha+\cos(\varphi(\alpha-1)+\phi)}
    {[\cos(\varphi\alpha)+\tau^\alpha(\Omega^2+(\nu+b)^2)^{\frac{\alpha}{2}}]^2+ \sin^2(\varphi\alpha)  },\\[3pt]
g=\frac{(\Omega^2+(\nu+b)^2)^{\frac{\alpha}{2}} \sin(\phi-\varphi)\tau^\alpha+\sin(\varphi(\alpha-1)+\phi)}
     {[\cos(\varphi\alpha)+\tau^\alpha(\Omega^2+(\nu+b)^2)^{\frac{\alpha}{2}}]^2+ \sin^2(\varphi\alpha)  },  \nonumber\\[3pt]
f_1=\frac{(\Omega^2+b^2)^{\frac{\alpha}{2}}\, \sin(\theta)\,\tau^\alpha-\sin(\theta(\alpha-1))}
     {[\cos(\theta\alpha)+\tau^\alpha\,(\Omega^2+b^2)^{\frac{\alpha}{2}}]^2+ \sin^2(\theta\alpha)  },  \nonumber\\[3pt]
g_1=\frac{(\Omega^2+b^2)^{\frac{\alpha}{2}}\, \cos(\theta)\,\tau^\alpha+\cos(\theta(\alpha-1))}
     {[\cos(\theta\alpha)+\tau^\alpha\,(\Omega^2+b^2)^{\frac{\alpha}{2}}]^2+ \sin^2(\theta\alpha)  },
      \nonumber\\[3pt]
m=\omega^2+\nu^2-\Omega^2
   +(\nu^2+\Omega^2)^{\frac{1}{2}}\, (\Omega^2+(\nu+b)^2)^{\frac{\alpha-1}{2}} \, \gamma \, f,  \nonumber\\[3pt]
n=2\Omega\nu+(\nu^2+\Omega^2)^{\frac{1}{2}}\, (\Omega^2+(\nu+b)^2)^{\frac{\alpha-1}{2}}\,  \gamma \, g,  \nonumber\\[3pt]
m_1=\omega^2-\Omega^2+\Omega\, (\Omega^2+b^2)^{\frac{\alpha-1}{2}} \, \gamma \, f_1,  \nonumber\\[3pt]
n_1=\Omega\, (\Omega^2+b^2)^{\frac{\alpha-1}{2}} \, \gamma \, g_1.
\end{eqnarray*}
It seems that the output amplitude $A$ not only depends on the periodic driving force, but also the multiplicative and additive noise. In the case $b=0$, these results recover the ones in \cite{Laas:2011}. As to (\ref{cor_as}), similarly, one can obtain
\begin{equation*}
H_{32}(-\rmi \Omega)=\chi'_1(\Omega)+\rmi \chi''_1(\Omega)
\end{equation*}
with
\begin{equation*}
\chi'_1(\Omega)=-\frac{a^2(m_1m-n_1n-a^2)}{(m_1m-n_1n-a^2)^2+(m_1n+n_1m)^2}
\end{equation*}
and
\begin{equation*}
\chi''_1(\Omega)=-\frac{a^2(m_1n+n_1m)}{(m_1m-n_1n-a^2)^2+(m_1n+n_1m)^2}.
\end{equation*}
Besides, we have
\begin{equation*}
A_1=A_0\frac{a^2}{((m_1m-n_1n-a^2)^2+(m_1n+n_1m)^2)^{\frac{1}{2}}}
\end{equation*}
and
\begin{equation*}
\Psi_1=\arctan\left(-\frac{m_1n+n_1m}{m_1m-n_1n-a^2}\right).
\end{equation*}
To sum up the previous formulae, $\langle x(t)\rangle_{as}$ in (\ref{fir_mon_as}) and its associated terms will be used in Section \ref{three}, while the formulae associated with $\langle z(t)x(t)\rangle_{as}$ in (\ref{cor_as}) will be used in Section \ref{four}.

\section{Results}\label{three}
\setcounter{equation}{0}
In this section, using the expression of amplitude $A$ in (\ref{A}), we will analyze the behavior of $A^2$ for different system parameters. In the first part, the behavior of $A^2$ versus the system parameters $\tau,~\Omega,~a$ are mainly discussed. In the second part, we study the influence of the exponential tempering on the critical memory exponent $\alpha_{cr}$ and the resonance regions where SR vs noise amplitude $a$ is possible.

\subsection{Output amplitude vs system parameters}
We analyze the behavior of the squared output amplitude $A^2$ versus the characteristic memory time $\tau$ and discuss the SR phenomenon, being understood in the wide sense, i.e.,
the non-monotonic behavior of the output signal or some functions of it (moments, autocorrelation functions, SNR) in respect to the noise and system parameters \cite{Gitterman:2005}.
Here, if the curve of the response function $A^2(\tau)$ is non-monotonic, we say that the SR phenomenon occurs. 

\begin{figure}[ht]
\begin{minipage}{0.31\linewidth}
  \centerline{\includegraphics[scale=0.37]{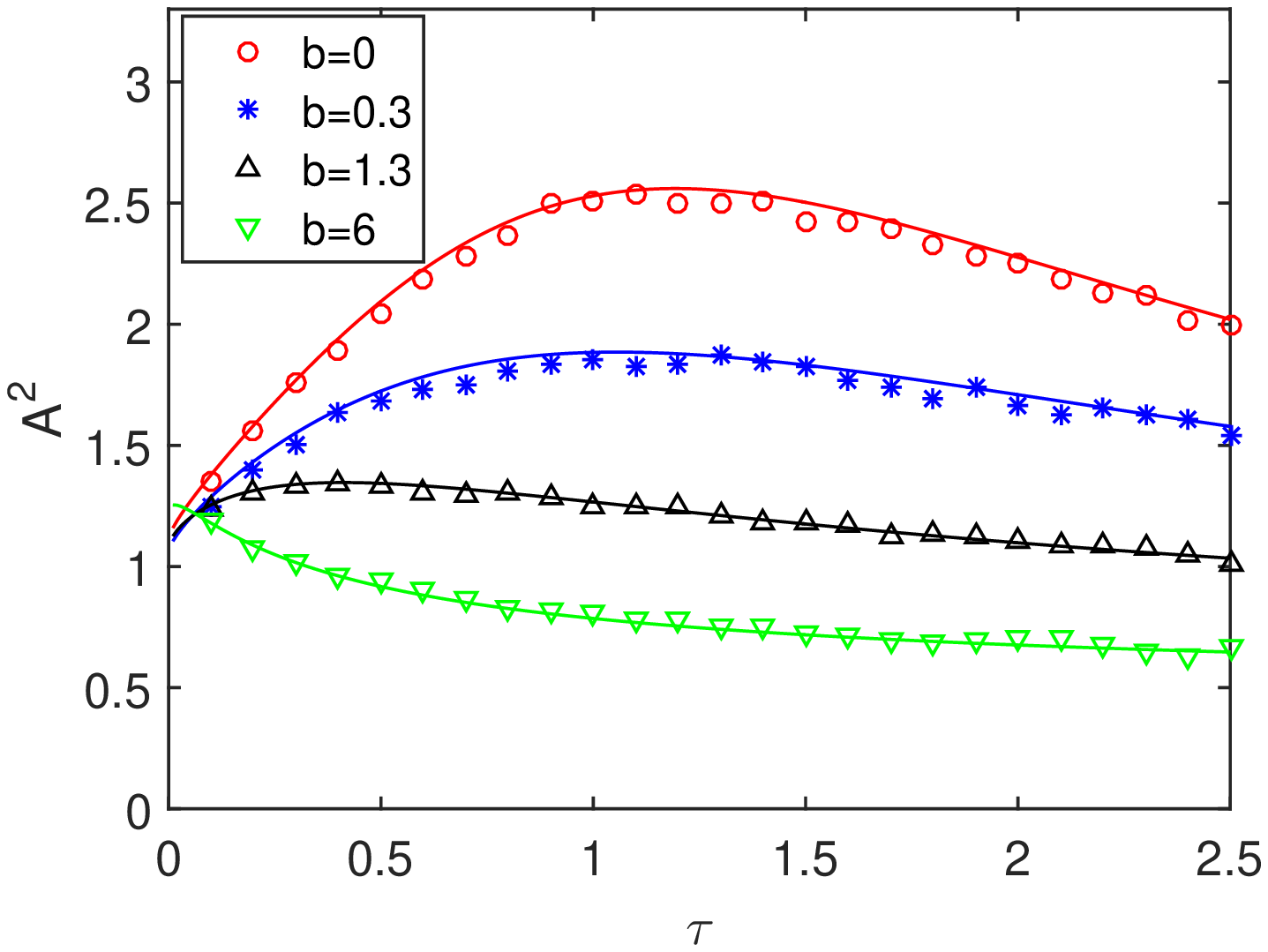}}
  \centerline{(a)}
\end{minipage}
\hfill
\begin{minipage}{0.31\linewidth}
  \centerline{\includegraphics[scale=0.37]{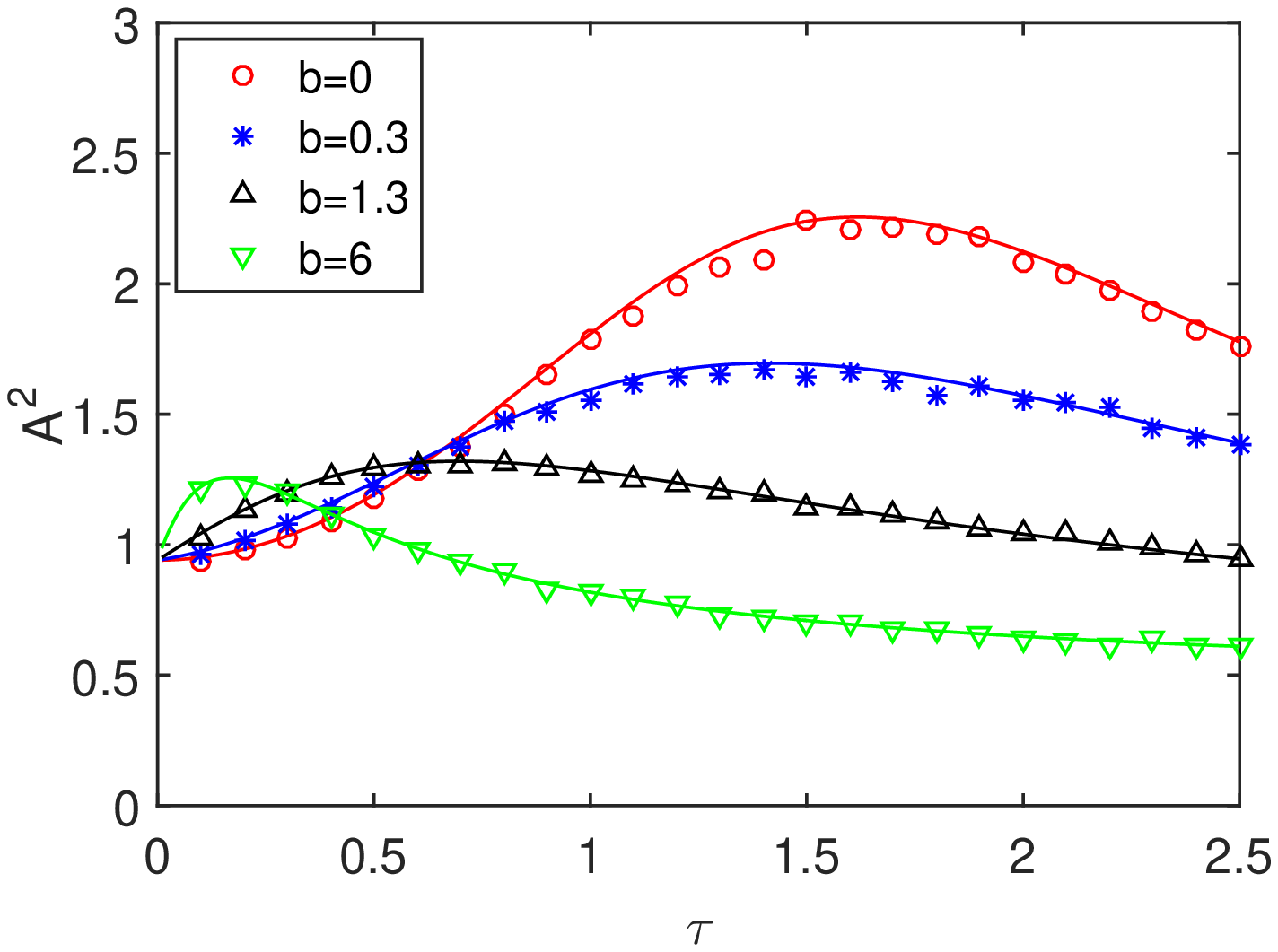}}
  \centerline{(b)}
\end{minipage}
\hfill
\begin{minipage}{0.31\linewidth}
  \centerline{\includegraphics[scale=0.37]{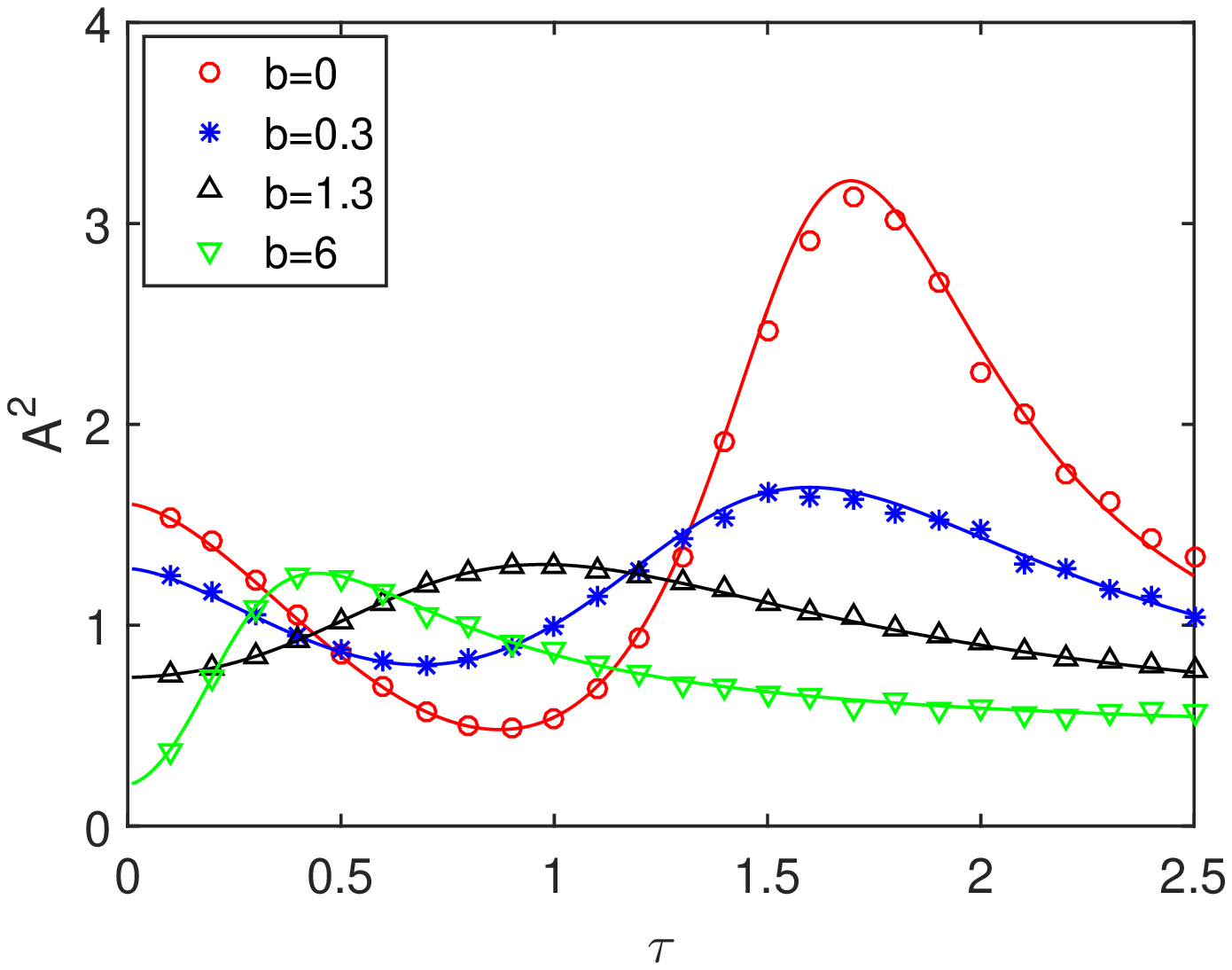}}
  \centerline{(c)}
\end{minipage}
  \caption{Squared output amplitude $A^2$ versus characteristic memory time $\tau$ with $\alpha=0.7$ for (a), $\alpha=1$ for (b), and $\alpha=1.6$ for (c).
Solid lines are the analytical results and the marks are the computer simulations with the time $T=200$ and the number of sampled trajectories $5000$. Other parameters: $\nu=0.1, A_0=\omega=1, a^2=0.3, \Omega=1$, and $\gamma=0.7$.
   }\label{Atau}
\end{figure}

Figure \ref{Atau} presents the squared output amplitude $A^2$ versus the characteristic memory time $\tau$ with different memory exponent $\alpha$ and tempering parameter $b$. It can be noted that for a fixed $\alpha$, with the increase of tempering parameter $b$, the resonance peak decreases and moves slightly to the left. In other words, with the increase of $b$, the appearance of the SR phenomenon needs shorter characteristic memory time,
 which, in some sense, can be explained that the increase of the tempering parameter $b$ reduces the memory effect of friction and then a shorter characteristic memory time $\tau$ can balance this effect. For fixed $\alpha<1$ (see Figure \ref{Atau}$(a)$), with the increase of $b$, the resonance peak keeps decreasing, until disappears. But for $\alpha>1$ (see Figure \ref{Atau}$(c)$), there still exists SR phenomenon although the tempering parameter $b$ is very large, which can even be several hundred. More precisely, there exist a peak and a valley in the curves of $A^2(\tau)$ for small $b$. However, there only exists a peak for large $b$ and the value of the peak does not change with the increase of $b$.
Next, we turn to the horizontal comparison with fixed tempering parameter $b$. When $b$ is small, with the increase of $\alpha$, it changes from having a peak to having a peak as well as a valley; for large $b$, the behavior of $A^2(\tau)$ is primarily monotonic (no SR) and then becomes non-monotonic (SR appears) with the increase of $\alpha$.

Another interesting issue is on the bona fide SR, which means that the squared output amplitude $A^2$ versus the driving frequency $\Omega$ exhibits a peak.
\begin{figure}[ht]
\begin{minipage}{0.31\linewidth}
  \centerline{\includegraphics[scale=0.37]{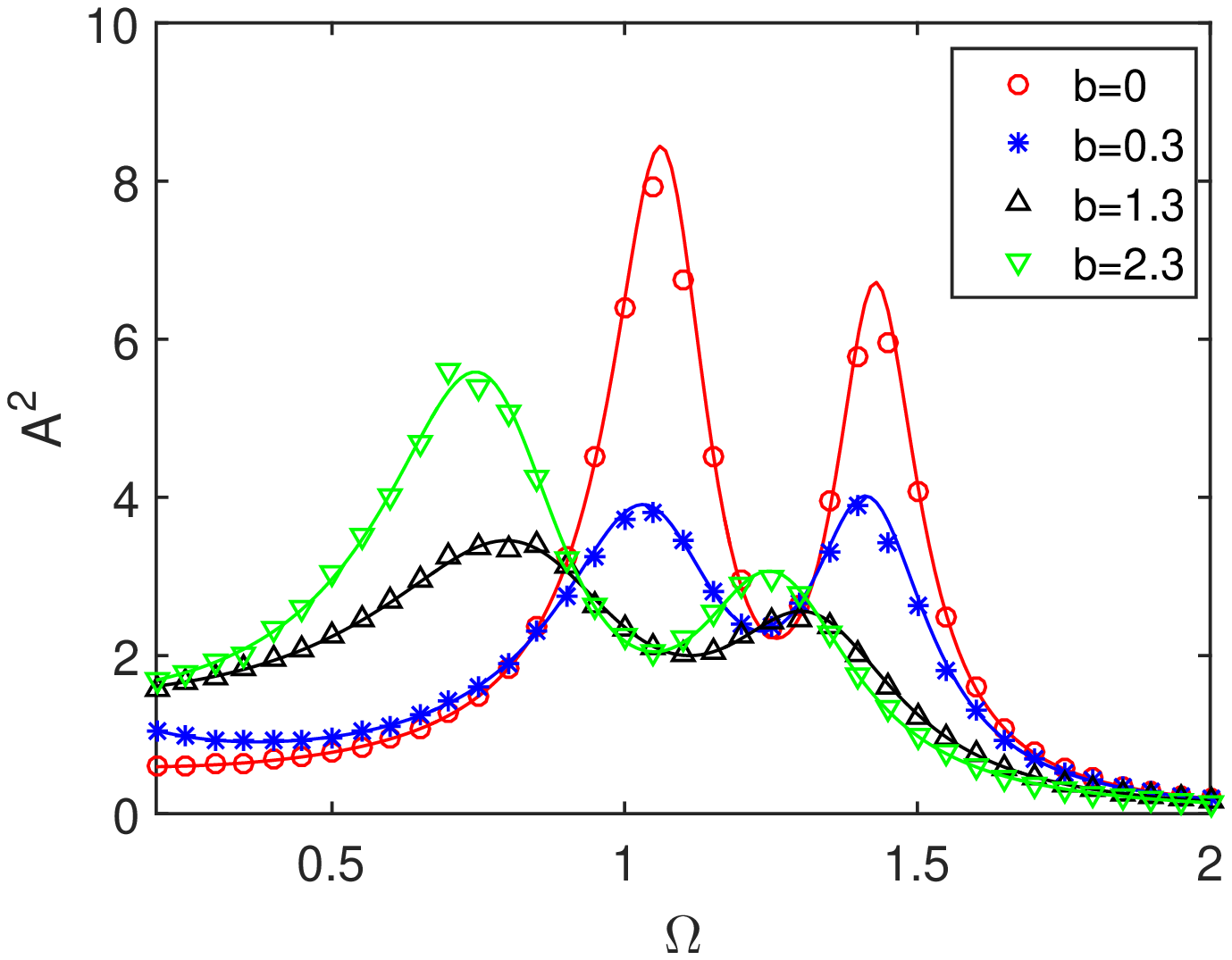}}
  \centerline{(a)}
\end{minipage}
\hfill
\begin{minipage}{0.31\linewidth}
  \centerline{\includegraphics[scale=0.37]{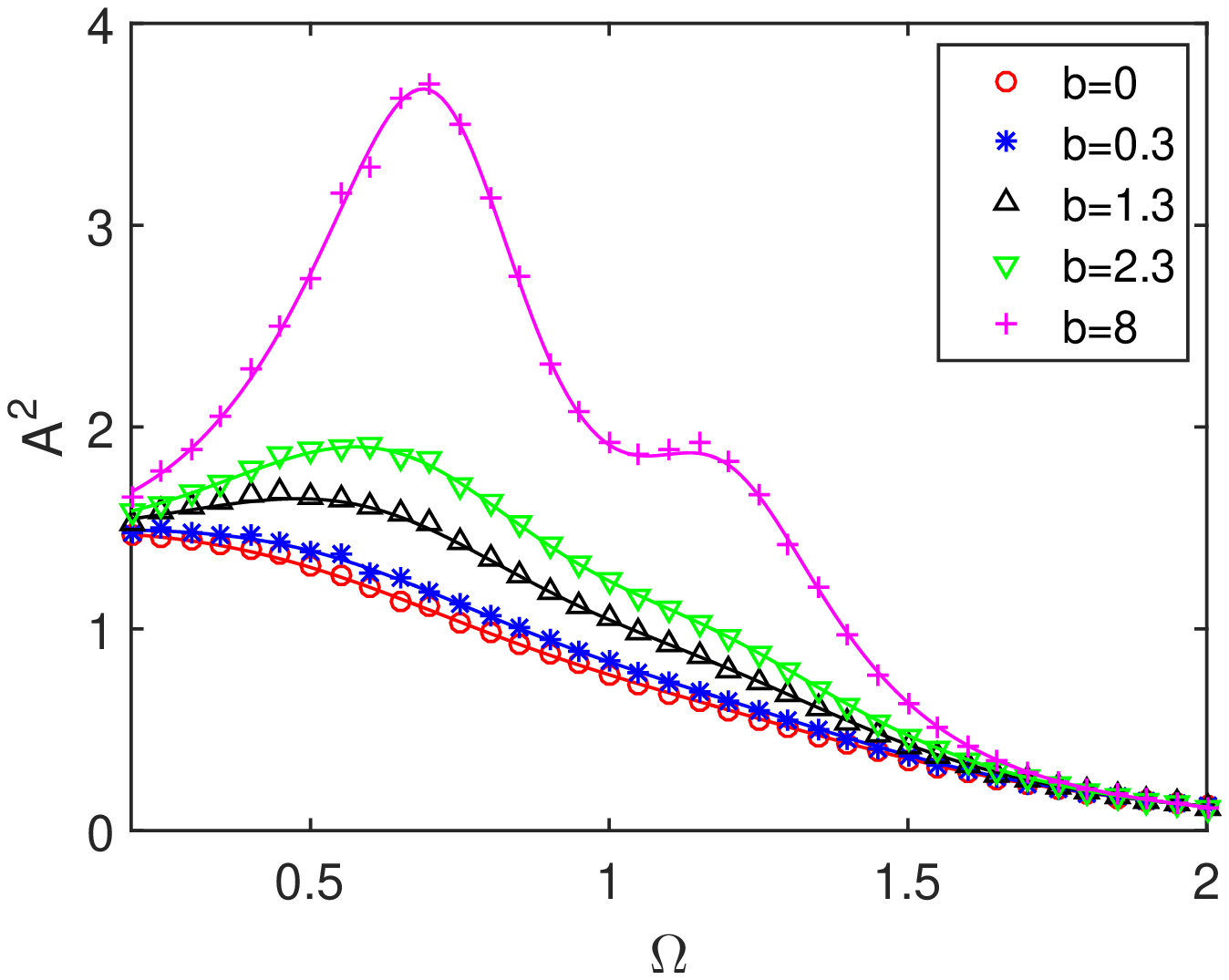}}
  \centerline{(b)}
\end{minipage}
\hfill
\begin{minipage}{0.31\linewidth}
  \centerline{\includegraphics[scale=0.37]{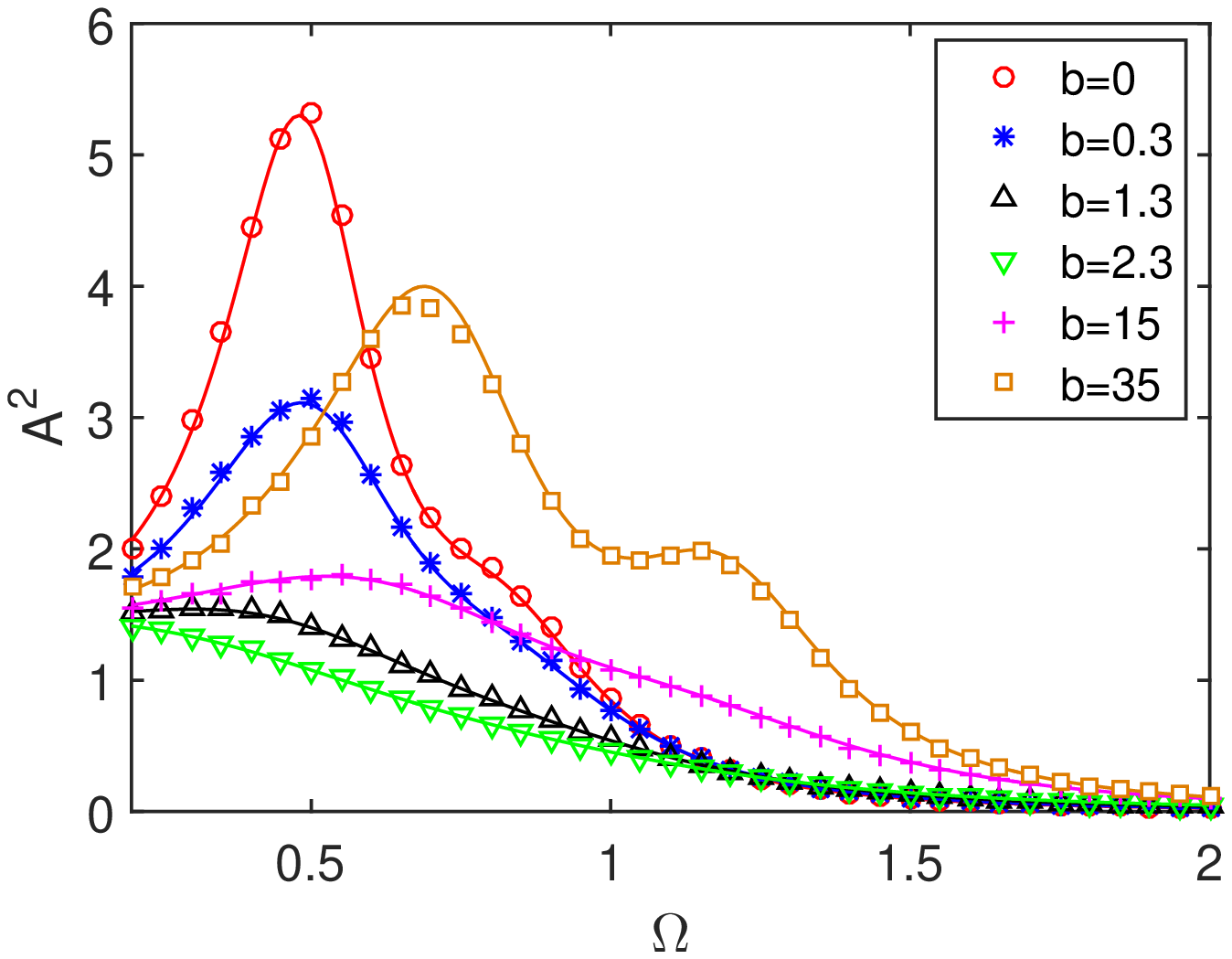}}
  \centerline{(c)}
\end{minipage}
\vfill
\begin{minipage}{0.31\linewidth}
  \centerline{\includegraphics[scale=0.37]{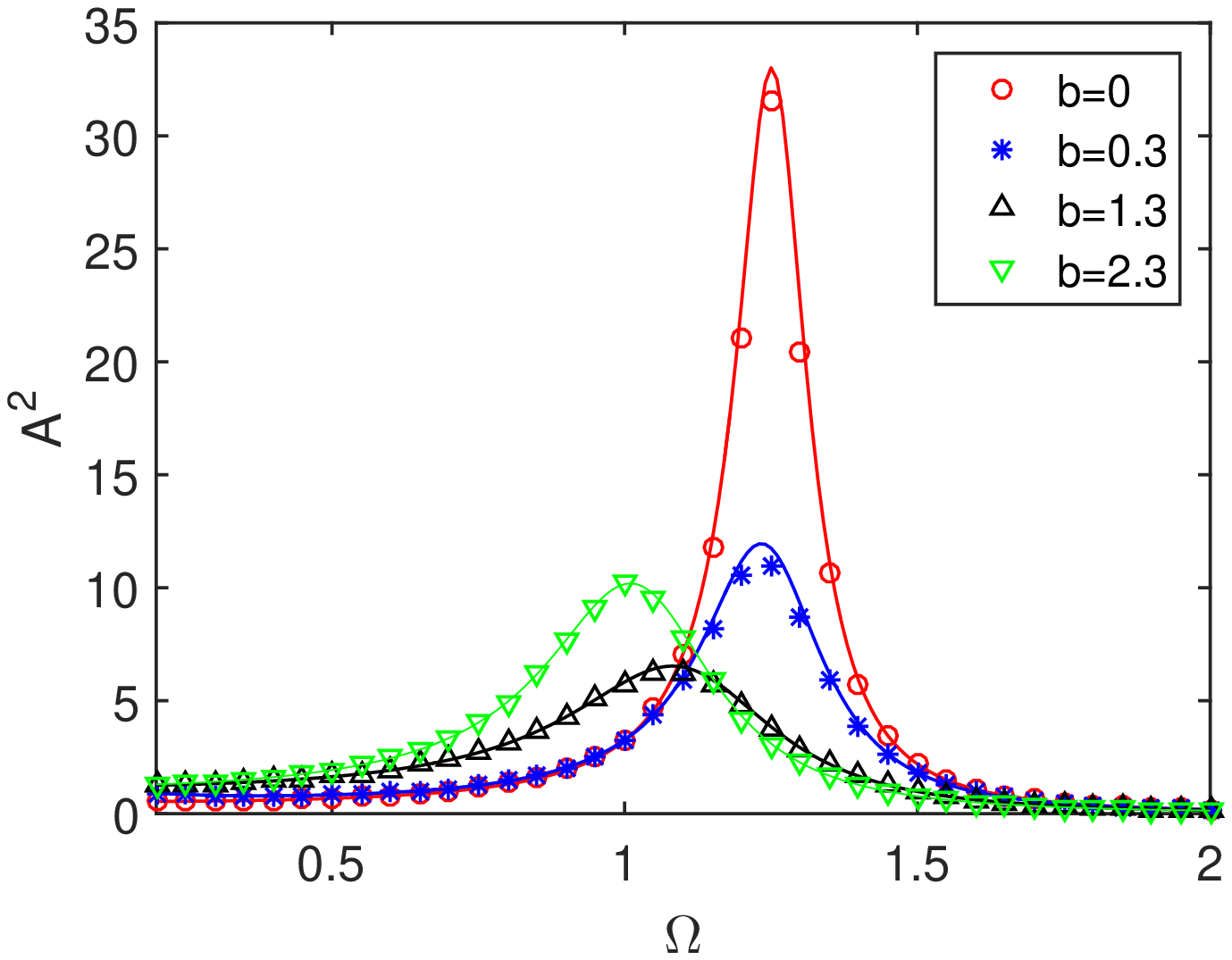}}
  \centerline{(d)}
\end{minipage}
\hfill
\begin{minipage}{0.31\linewidth}
  \centerline{\includegraphics[scale=0.37]{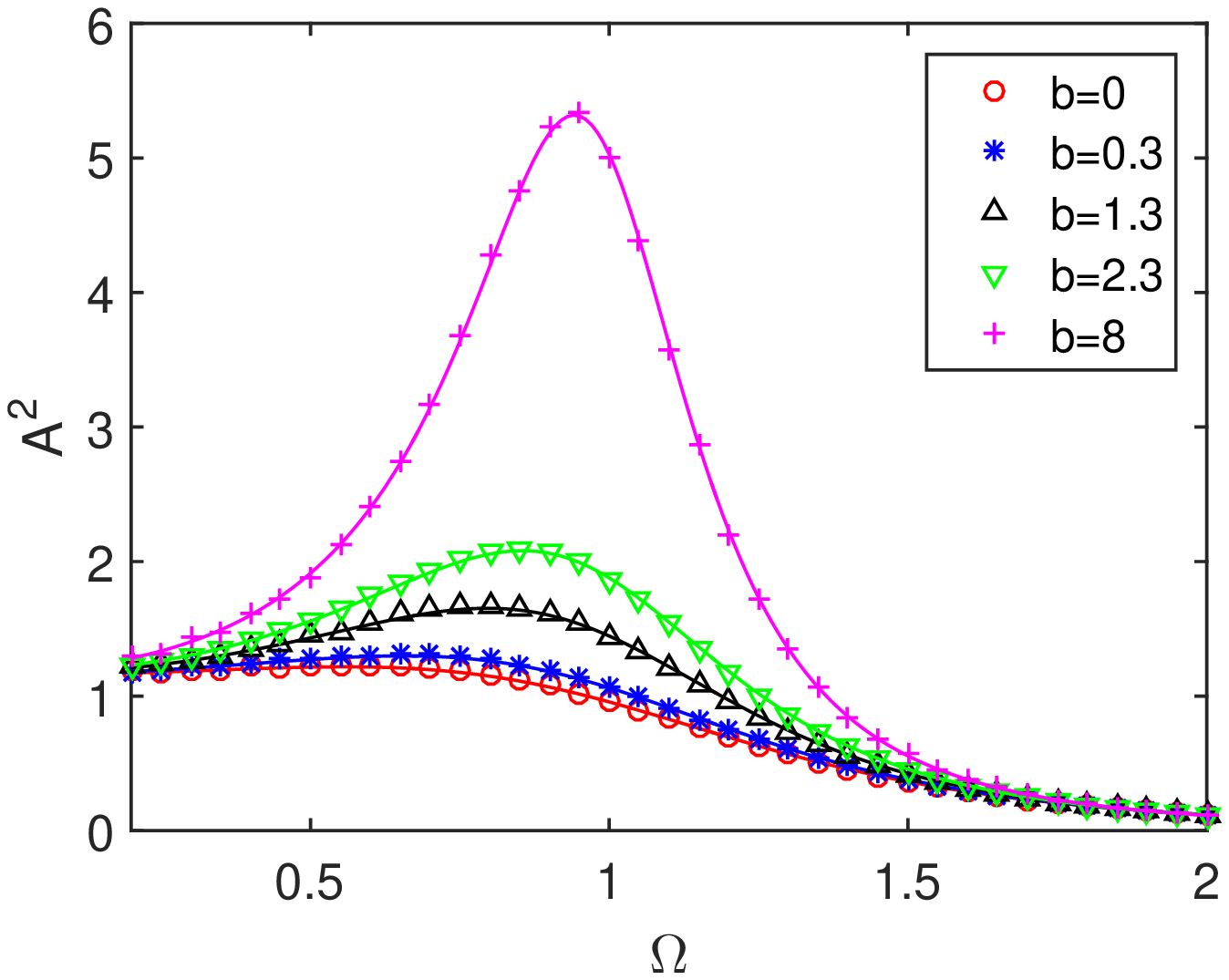}}
  \centerline{(e)}
\end{minipage}
\hfill
\begin{minipage}{0.31\linewidth}
  \centerline{\includegraphics[scale=0.37]{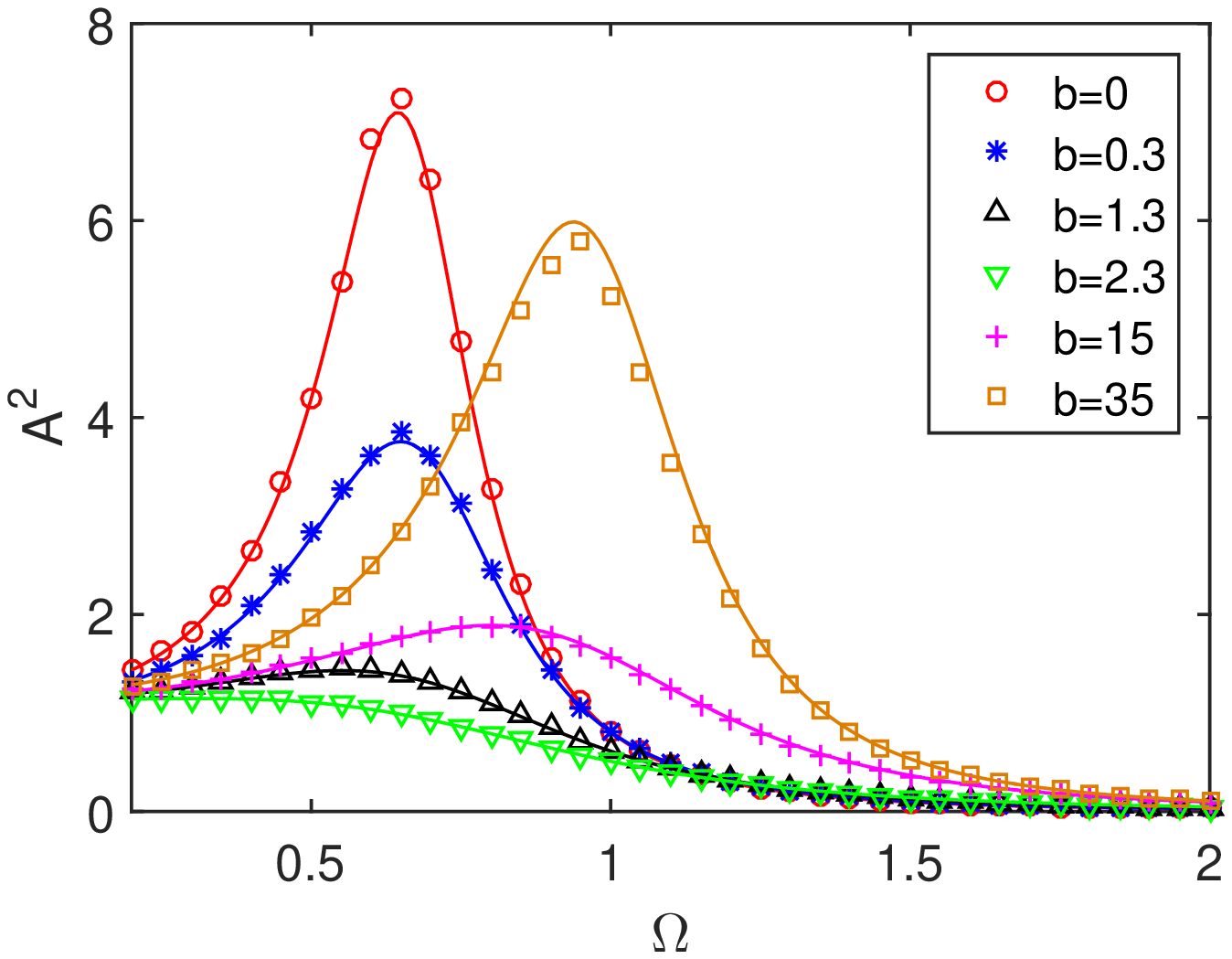}}
  \centerline{(f)}
\end{minipage}
  \caption{Squared output amplitude $A^2$ versus the angular frequency $\Omega$ of the harmonic driving force. Solid lines are the analytical results and the marks are the computer simulations with the time $T=100$ and the number of sampled trajectories $5000$. Parameter values: $A_0=\omega=1, a^2=0.2, \tau=0.2, \gamma=1$; (a): $\alpha=0.2, \nu=0.1$; (b): $\alpha=1, \nu=0.1$; (c): $\alpha=1.6, \nu=0.1$; (d): $\alpha=0.2, \nu=1$; (e): $\alpha=1, \nu=1$; (f): $\alpha=1.6, \nu=1$. }\label{AOmg}
\end{figure}
In Figure \ref{AOmg}, we show the dynamical behavior of $A^2(\Omega)$ with respect to the tempering  parameter $b$, memory exponent $\alpha$, and the switching rate $\nu$, which uncovers the bona fide SR phenomenon. Figure \ref{AOmg}(a) ($\alpha=0.2$) displays a double-peak phenomenon. Besides, with the increase of the tempering parameter $b$, the resonance peak is primarily restrained and then increases, and moves slightly to the left. Figure \ref{AOmg}(b) ($\alpha=1$) shows that with the increase of the tempering parameter $b$, $A^2(\Omega)$ is primarily monotonous ($b<0.3$), then a resonance peak appears and increases gradually. Finally, with the sustained increase of $b$, a double-peak phenomenon appears. In Figure \ref{AOmg}(c) ($\alpha=1.6$), with the increase of the tempering parameter $b$, the resonance peak is primarily restrained ($A^2(\Omega)$ is monotonous when $b=2.3$), then increases and, finally, a double-peak phenomenon appears.

The above phenomena imply that the exponentially tempered Mittag-Leffler dissipation memory kernel can be taken as the exponent-form one for a sufficiently large tempering parameter $b$ (obviously, for getting the similar dynamics, in the case of $\alpha>1$, the tempering parameter $b$ needs to be larger than the case of $\alpha<1$). Thus, with the increase of the tempering parameter $b$, the dynamic behaviors of $A^2(\Omega)$ in Figure \ref{AOmg}(a) and Figure \ref{AOmg}(c) are finally similar to the case $\alpha=1$ in Figure \ref{AOmg}(b). More precisely, the output amplitude finally increases with the increase of $b$ and a double-peak phenomenon appears; see the three representations $b=2.3$, $b=8$, $b=35$ in Figure \ref{AOmg}(a)(b)(c) respectively.
And the positions of the peaks in Figure \ref{AOmg}(a)(b)(c) are almost the same for respective large $b$.

When the noise switching rate $\nu$ increases, the double-peak phenomena can not be observed (see Figure \ref{AOmg}(d)(e)(f)). The reason is that the dynamic system (\ref{SR}) tends to deterministic system with harmonic potential $V(x)=\omega^2\frac{x^2}{2}$ \cite{Soika:2010}. But the movement and the raise or suppression of the resonance peak are similar to the case of small $\nu$. Besides, with the increase of the noise switching rate $\nu$, the driving frequency $\Omega$ needed for the maximum peak becomes larger.
\begin{figure}[ht]
\begin{minipage}{0.67\linewidth}
  \centerline{\includegraphics[scale=0.4]{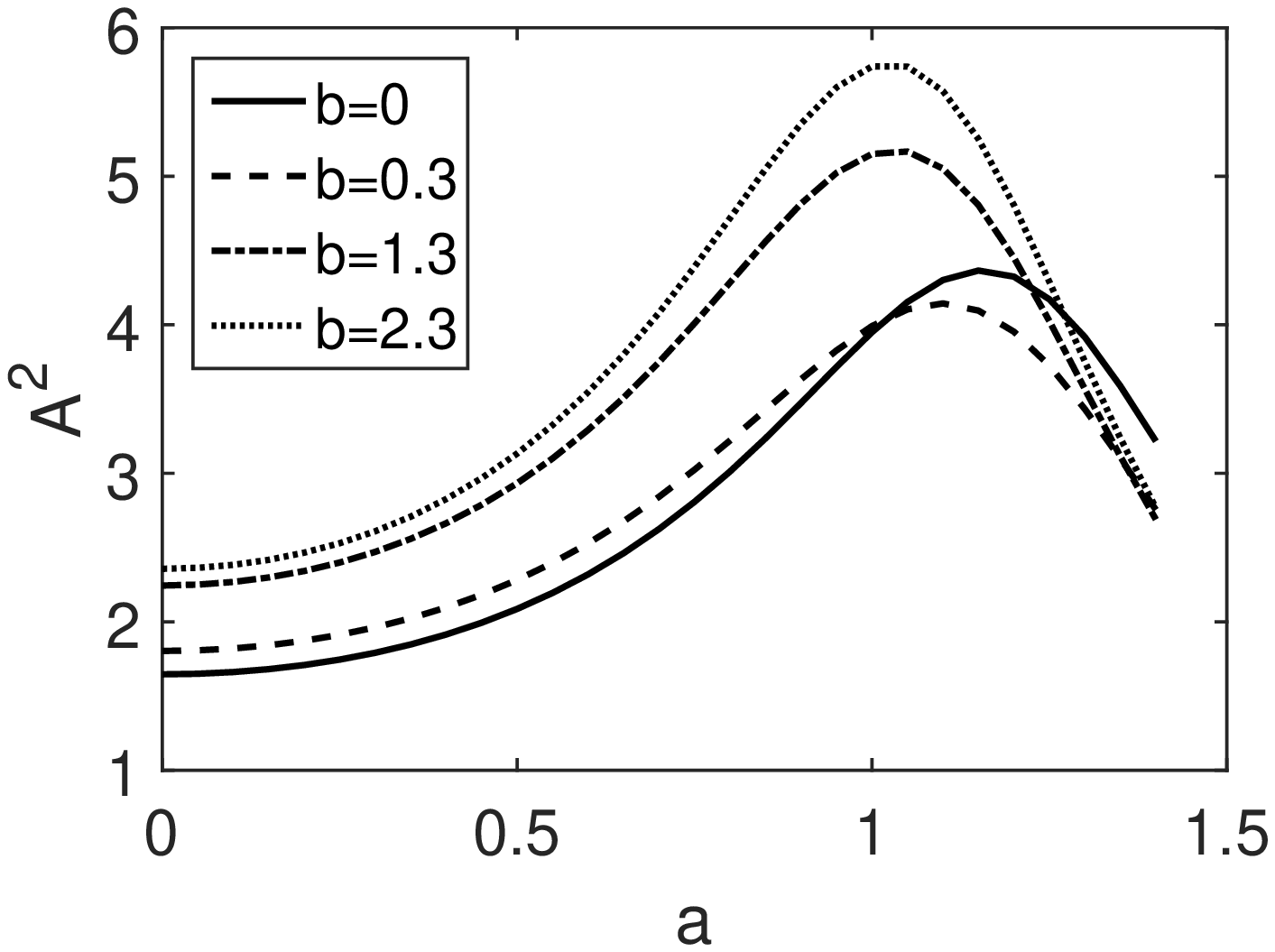}}
  \centerline{(a)}
\end{minipage}
\hfill
\begin{minipage}{0.33\linewidth}
  \centerline{\includegraphics[scale=0.4]{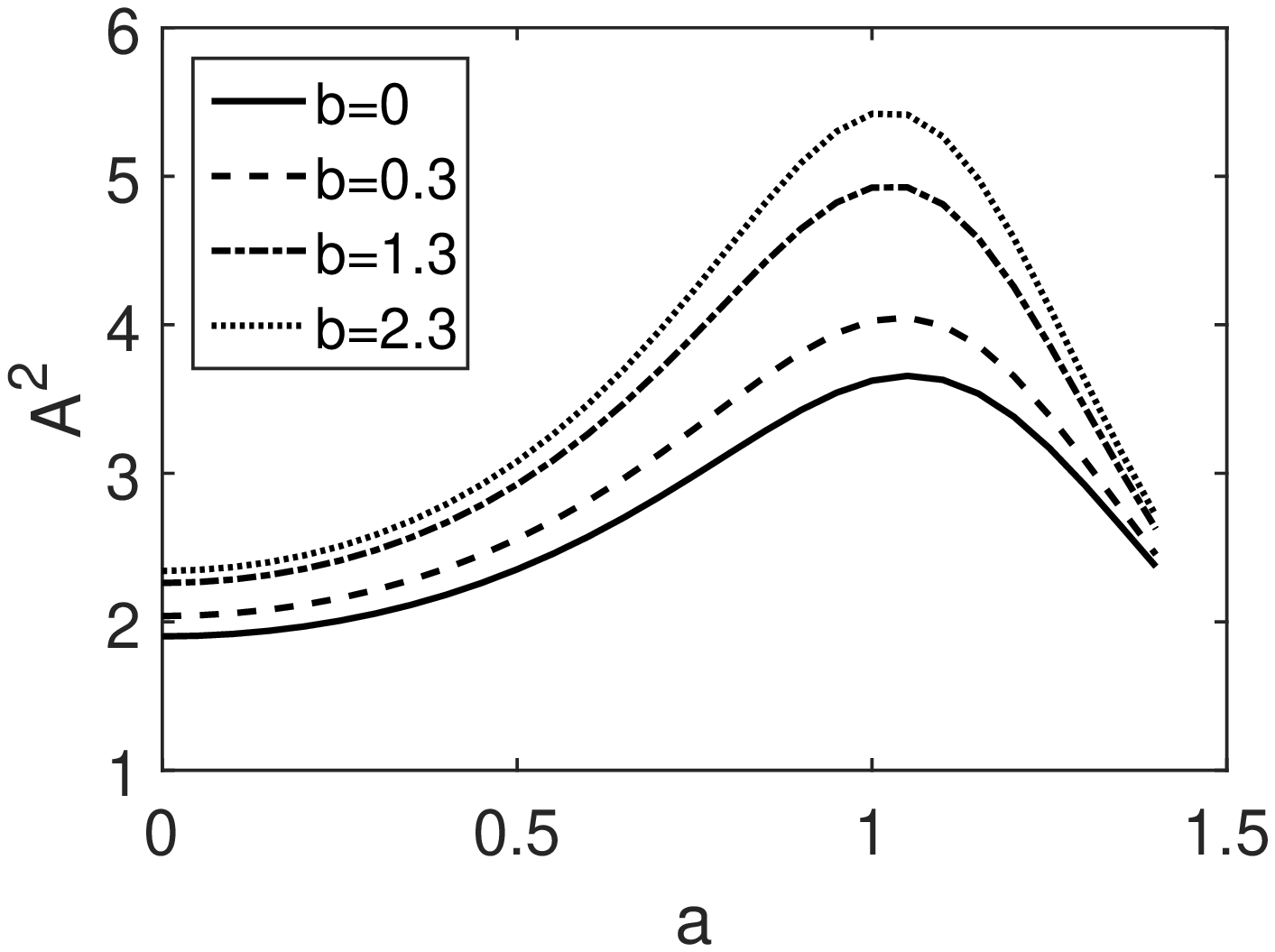}}
  \centerline{(b)}
\end{minipage}
\vfill
\begin{minipage}{0.67\linewidth}
  \centerline{\includegraphics[scale=0.4]{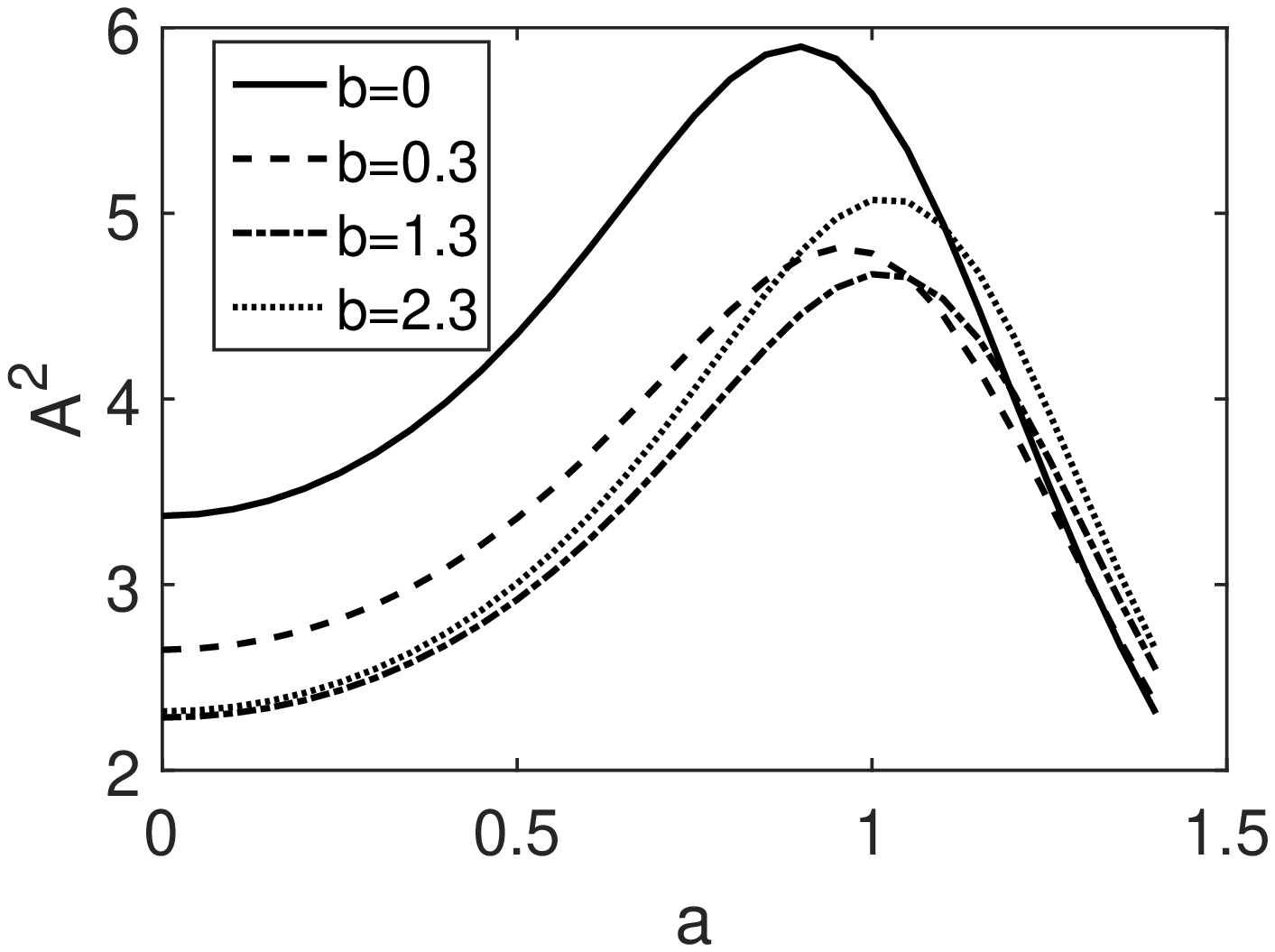}}
  \centerline{(c)}
\end{minipage}
\hfill
\begin{minipage}{0.33\linewidth}
  \centerline{\includegraphics[scale=0.4]{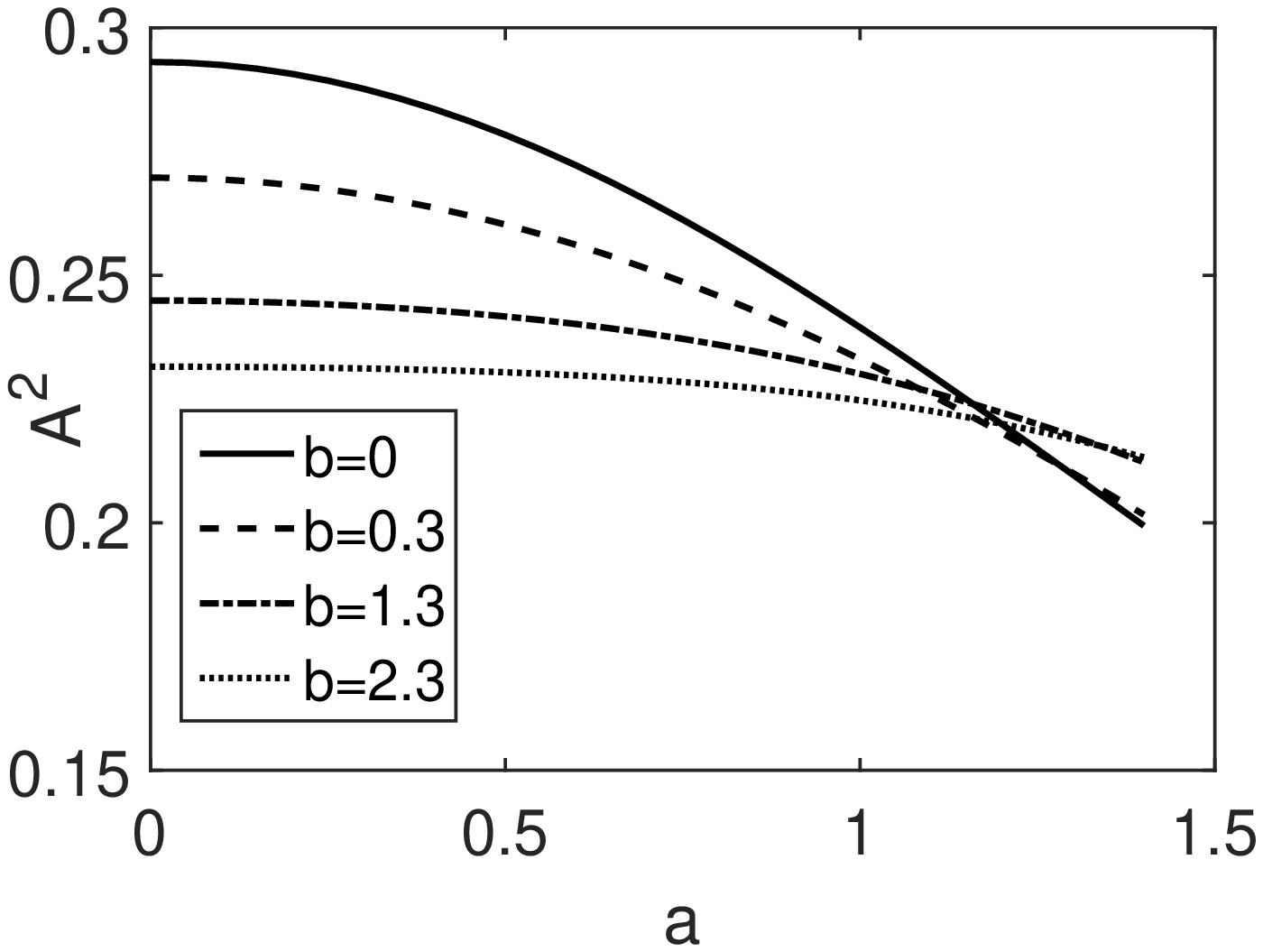}}
  \centerline{(d)}
\end{minipage}
  \caption{Squared output amplitude $A^2$ versus the dichotomous noise amplitude $a$. Parameter values: $\nu=1, A_0=\omega=1$; (a): $\Omega=0.6$, $\tau=0.8$, $\gamma=0.3$, $\alpha=0.4$; (b): $\Omega=0.6$, $\tau=0.8$, $\gamma=0.3$, $\alpha=1$; (c): $\Omega=0.6$, $\tau=0.8$, $\gamma=0.3$, $\alpha=1.6$; (d): $\Omega=1.8$, $\tau=0.65$, $\gamma=0.6$, $\alpha=1.6$. We emphasize that these parameters satisfy the stability condition (\ref{stability}). In addition, if we keep the  parameters as the same as above except $\nu=0.1$, the SR phenomena of $A(a)$ are  similar.}\label{Aa}
\end{figure}
We further examine the dependence of the squared response function $A^2$ on the amplitude $a$ of the multiplicative symmetric dichotomous noise. In Figure \ref{Aa}, we display the curves of $A^2(a)$ with respect to the tempering parameter $b$ and the memory exponent $\alpha$. There exist peaks in Figure \ref{Aa}(a)(b)(c), which means that the typical SR phenomena appear \cite{Mankin:2008}. From Figure \ref{Aa}(a) ($\alpha=0.4$), it can be noted that the increase of the tempering parameter $b$ first suppresses the resonance peak, and then enhances the SR; the position of the peak moves slightly to the left, then remains unchanged. Figure \ref{Aa}(b) ($\alpha=1$) shows that the resonance peak increases with the increase of the tempering parameter $b$, and the position of the peak remains unchanged. In Figure \ref{Aa}(c) ($\alpha=1.6$), the resonance peak is first restrained and then increases with the increase of the tempering parameter $b$. But contrary to Figure \ref{Aa}(a), the position of the peak moves slightly to the right, then remains unchanged. The final positions of the resonance peaks are almost the same in Figure \ref{Aa}(a)(b)(c), the phenomenon of which is also detected in Figure \ref{AOmg} and explained there.
Besides, the parameter values in Figure \ref{Aa}(d) are consistent with those in Figure \ref{gama_alp_have_v}(h), in which the reason of no SR phenomenon is uncovered, i.e., the frictional coefficient $\gamma$ is not small enough.
 This implies that the existence of SR strongly depends on the system parameter.

\subsection{SR regions in parameter space $(\gamma, \alpha)$}
\begin{figure}[ht]
\flushright
\begin{minipage}{0.65\linewidth}
  \centerline{\includegraphics[scale=0.38]{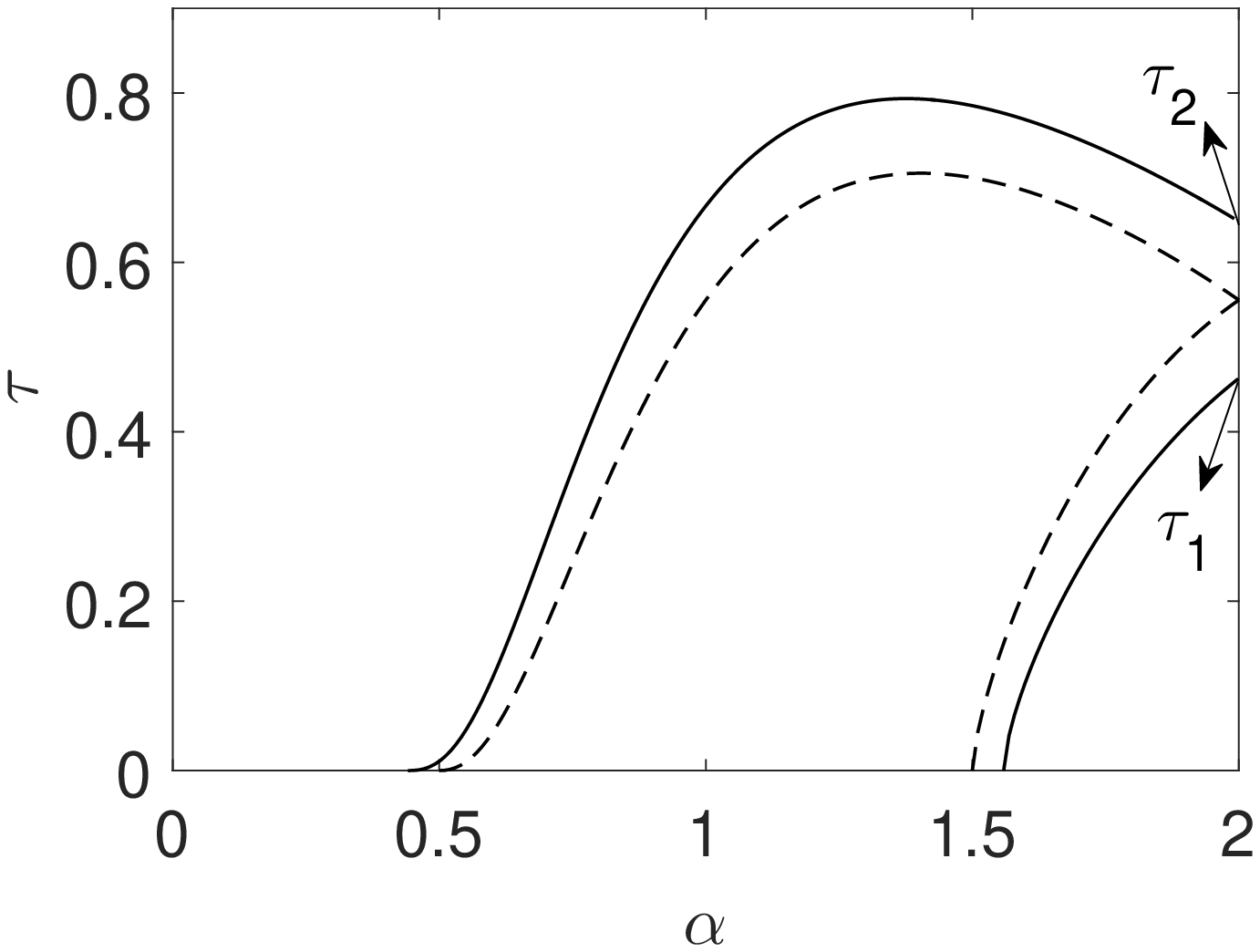}}
  \centerline{(a)}
\end{minipage}
\hfill
\begin{minipage}{0.31\linewidth}
  \centerline{\includegraphics[scale=0.38]{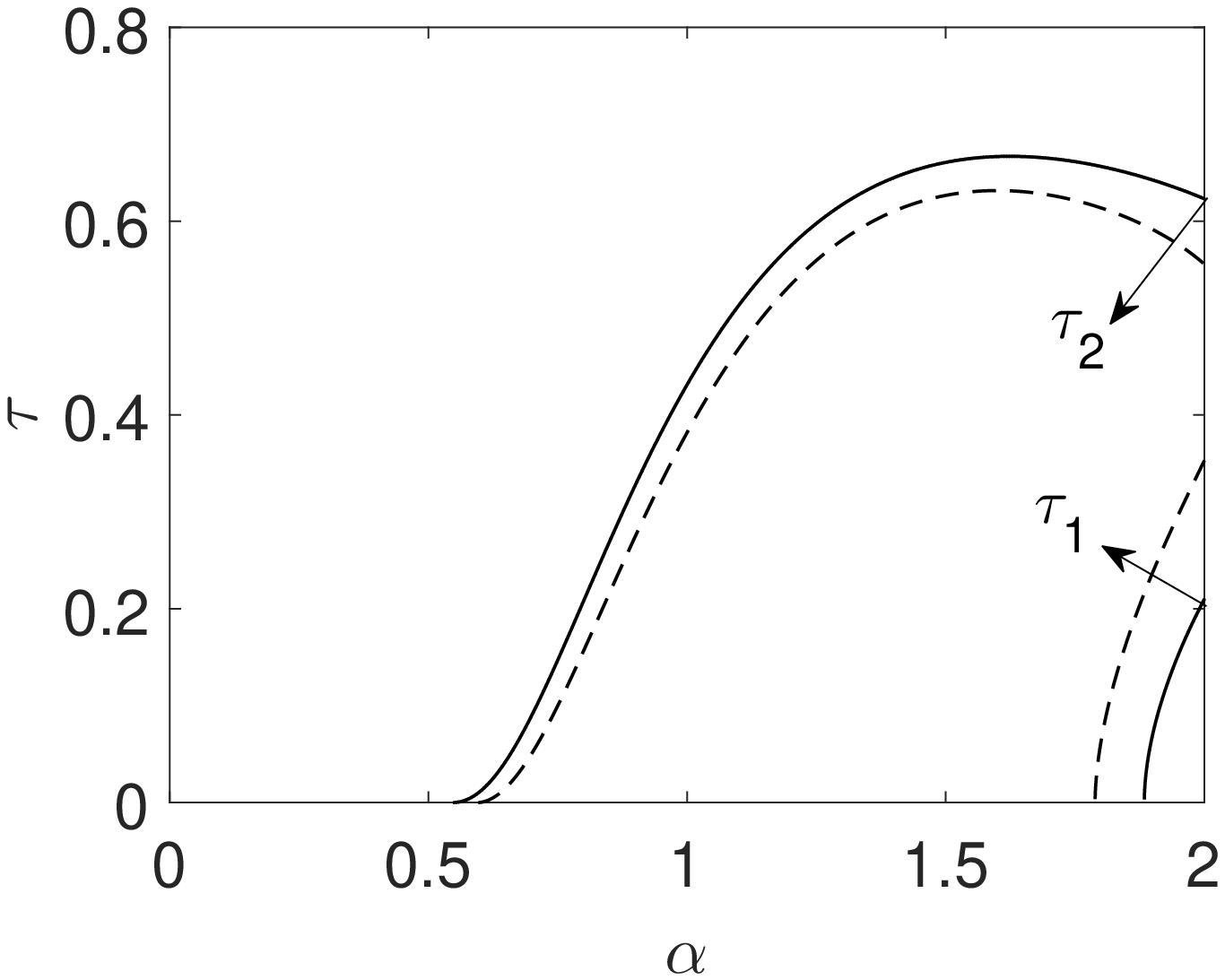}}
  \centerline{(b)}
\end{minipage}
  \caption{Characteristic memory time $\tau$ versus memory exponent $\alpha$. Parameter values: (a): $\nu=0$, $\Omega=1.8$, $b=0.3$ (solid line), $b=0$ (dashed line); (b): $\nu=1$, $\Omega=1.8$, $b=0.3$ (solid line), $b=0$ (dashed line).  }\label{tau_alp}
\end{figure}

In this part, the relation between the friction constant $\gamma$ and the memory exponent $\alpha$ ($0<\alpha<2$), as well as the influence of the tempering parameter $b$, for the emergence of SR versus noise amplitude $a$ is studied. 
From (\ref{A}), we know that the response $A(a)$ reaches its maximum at $a^2_{max}=m_1m-n_1n$. Then because of the stability condition (\ref{stability}) $0<a<a_{cr}$, the non-monotonic behavior of $A(a)$ in the region $[0,a_{cr}]$ can be guaranteed if $0<m_1m-n_1n<a^2_{cr}$.
One of our tasks is to discuss the adiabatic multiplicative noise, i.e., $\nu\rightarrow0$; Figure \ref{tau_alp}(a) is related to this case. In this case, the boundary lines between the regions where SR (versus $a$) occurs or not in the parameter space $(\gamma, \alpha)$ is as follows
\begin{equation}
\gamma_{1,2}(\alpha)=\frac{G_1}{G_2\pm G_3},
\end{equation}
where $0<\alpha<2$ and
\begin{eqnarray}
G_1=(\omega^2-\Omega^2)\, \big([\cos(\theta\alpha)+\tau^\alpha\, (\Omega^2+b^2)^{\alpha/2}]^2+\sin^2(\theta\alpha)\big), \nonumber\\[3pt]
G_2=-(\Omega^2+b^2)^{(2\alpha-1)/2}\, \Omega\, \tau^\alpha\, \sin(\theta)+(\Omega^2+b^2)^{(\alpha-1)/2}\, \Omega\, \sin(\theta\alpha-\theta), \nonumber\\[3pt]
G_3=(\Omega^2+b^2)^{(2\alpha-1)/2}\, \Omega\, \tau^\alpha\, \cos(\theta)+(\Omega^2+b^2)^{(\alpha-1)/2}\, \Omega\, \cos(\theta\alpha-\theta). \nonumber
\end{eqnarray}
After reasonably choosing some parameters, there will be a critical memory exponent $\alpha_{cr}$, where a sharp transition of the dynamical behaviors of the system happens; at $\alpha_{cr}$, one of the boundaries $\gamma_{1,2}(\alpha)$ between the resonance and non-resonance regions tends to infinity. Taking $G_2\pm G_3=0$, we obtain that $\alpha_{cr}$ satisfies the following expressions
\begin{equation}\label{alpha1}
-(\Omega^2+b^2)^{\alpha_{cr}/2}\tau^{\alpha_{cr}}=
\frac{\textrm{sin}(\frac{\pi}{4}-(1-\alpha_{cr})\theta)}{\textrm{sin}(\frac{\pi}{4}-\theta)}, \qquad 0<\alpha_{cr}<2,
\end{equation}
\begin{equation}\label{alpha2}
-(\Omega^2+b^2)^{\alpha_{cr}/2}\tau^{\alpha_{cr}}=
\frac{\textrm{sin}(\frac{\pi}{4}+(1-\alpha_{cr})\theta)}{\textrm{sin}(\frac{\pi}{4}+\theta)}, \qquad  1<\alpha_{cr}<2,
\end{equation}
where $\theta\in(0, \frac{\pi}{2})$. It is obvious that the value of $\alpha_{cr}$ depends on the driving frequency $\Omega$, tempering parameter $b$, and characteristic memory time $\tau$. Moreover, if $b\geq\Omega$, i.e., $\theta\in(0, \frac{\pi}{4}]$, there is no root for (\ref{alpha1}) and (\ref{alpha2}), which means that no $\alpha_{cr}$ induces $\gamma_{1,2}(\alpha_{cr})\rightarrow\infty$. If $b<\Omega$, i.e.,  $\theta\in(\frac{\pi}{4}, \frac{\pi}{2})$, from (\ref{alpha1}), we have
\begin{enumerate}
  \item $\alpha_{cr}\geq\alpha_{crmin}=1-\frac{\pi}{4\theta}\in(0, 1/2]$,
       and $\alpha_{crmin}$ decreases as the tempering parameter $b$ increases;
  \item If $b=0$, one has $\alpha_{cr}\geq\frac{1}{2}$, which recovers the result in \cite{Laas:2011};
  \item If $\tau\rightarrow0$, (\ref{TML}) can be treated as the tempered power law memory kernel. In this case, (\ref{alpha1}) becomes $\alpha_{cr}=1-\frac{\pi}{4\theta}$. Taking $b=0$ again, one has $\alpha_{cr}=\frac{1}{2}$, which recovers the conclusion in \cite{Soika:2010}.
\end{enumerate}
   Besides, from (\ref{alpha2}), we have
   \begin{enumerate}
     \item $2>\alpha_{cr}\geq\alpha_{crmin}=1+\frac{\pi}{4\theta}\in[3/2, 2)$, and $\alpha_{crmin}$ increases as the tempering parameter $b$ increases;
     \item If $b=0$, one has $\alpha_{cr}\geq\frac{3}{2}$;
     \item If $\tau\rightarrow0$, (\ref{TML}) can be treated as the tempered power law memory kernel and it is nonsingular at origin. In this case, (\ref{alpha2}) becomes $\alpha_{cr}=1+\frac{\pi}{4\theta}$. Taking $b=0$ again, one has $\alpha_{cr}=\frac{3}{2}$.
   \end{enumerate}

Then, we consider the number of the critical memory exponents $\alpha_{cr}$ for any given  characteristic memory time $\tau$. The relationship between $\tau$ and $\alpha$ is shown in Figure \ref{tau_alp}(a), and it can be found that
\begin{enumerate}
  \item If $0\leq\tau\leq\tau_1$, there exist $\alpha_{cr1}$ and $\alpha_{cr2}$, s.t., $\gamma_1(\alpha_{cr1})\rightarrow\infty$ and $\gamma_2(\alpha_{cr2})\rightarrow\infty$;
  \item If $\tau_2\leq\tau<\tau_{max}$, there exist $\alpha_{cr11}$ and $\alpha_{cr12}$, s.t., $\gamma_1(\alpha_{cr11})\rightarrow\infty$ and $\gamma_1(\alpha_{cr12})\rightarrow\infty$;
  \item If $\tau_1<\tau<\tau_2$, or $\tau=\tau_{max}$, there exists $\alpha_{cr1}$, s.t., $\gamma_1(\alpha_{cr1})\rightarrow\infty$;
  \item If $\tau>\tau_{max}$, there is no $\alpha_{cr}$, s.t., $\gamma_1(\alpha_{cr})\rightarrow\infty$ or $\gamma_2(\alpha_{cr})\rightarrow\infty$.
\end{enumerate}
   Here $\tau_1=\left(\frac{\Omega-b}{(\Omega+b)(\Omega^2+b^2)}\right)^{1/2}$, $\tau_2=\left(\frac{\Omega+b}{(\Omega-b)(\Omega^2+b^2)}\right)^{1/2}$ and $\tau$ reaches its maximum value $\tau_{max}$ when $\alpha=25.4357\, \cos(0.3735\frac{b}{\Omega}-0.0414)-24.0431$ by the fitting method. In addition, for $b=0$, one has $\tau_1=\tau_2=\frac{1}{\Omega}$; see the dashed lines in Figure \ref{tau_alp}(a).

\begin{figure}[ht]
\begin{minipage}{0.31\linewidth}
  \centerline{\includegraphics[scale=0.6]{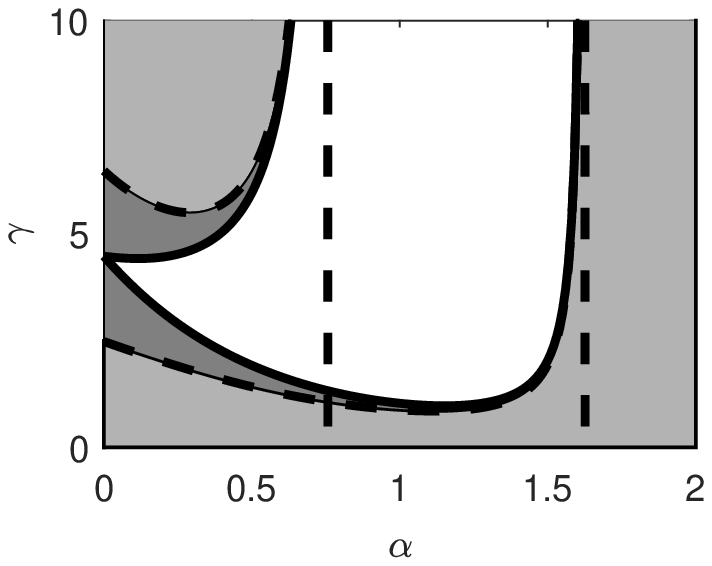}}
  \centerline{(a)}
\end{minipage}
\hfill
\begin{minipage}{0.31\linewidth}
  \centerline{\includegraphics[scale=0.6]{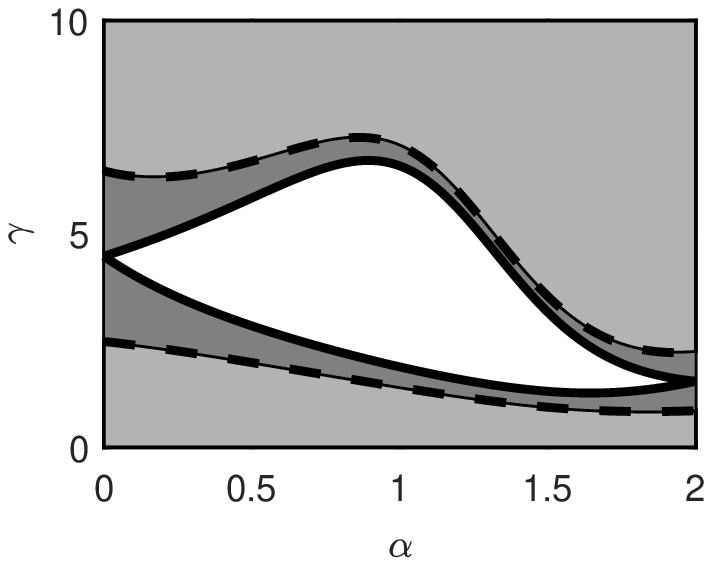}}
  \centerline{(b)}
\end{minipage}
\hfill
\begin{minipage}{0.31\linewidth}
  \centerline{\includegraphics[scale=0.6]{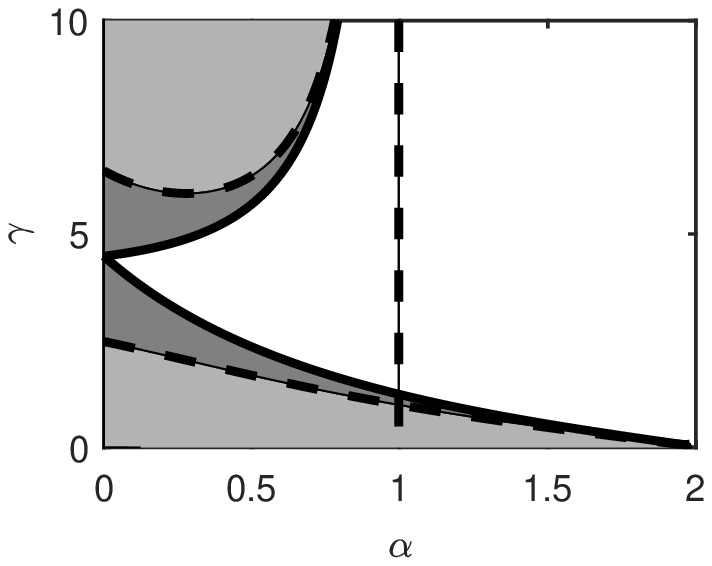}}
  \centerline{(c)}
\end{minipage}
\vfill
\begin{minipage}{0.31\linewidth}
  \centerline{\includegraphics[scale=0.6]{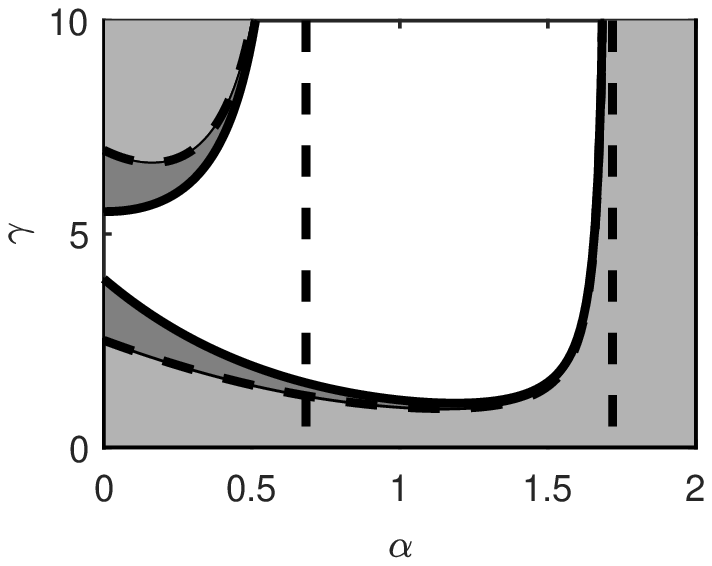}}
  \centerline{(d)}
\end{minipage}
\hfill
\begin{minipage}{0.31\linewidth}
  \centerline{\includegraphics[scale=0.6]{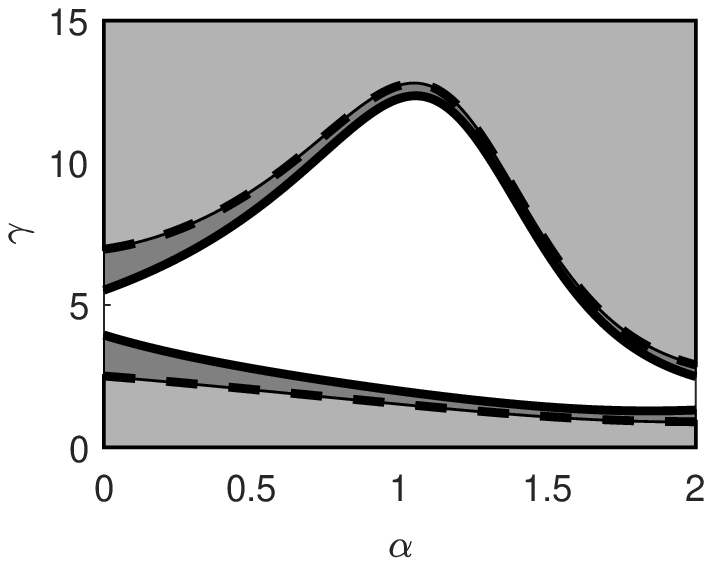}}
  \centerline{(e)}
\end{minipage}
\hfill
\begin{minipage}{0.31\linewidth}
  \centerline{\includegraphics[scale=0.6]{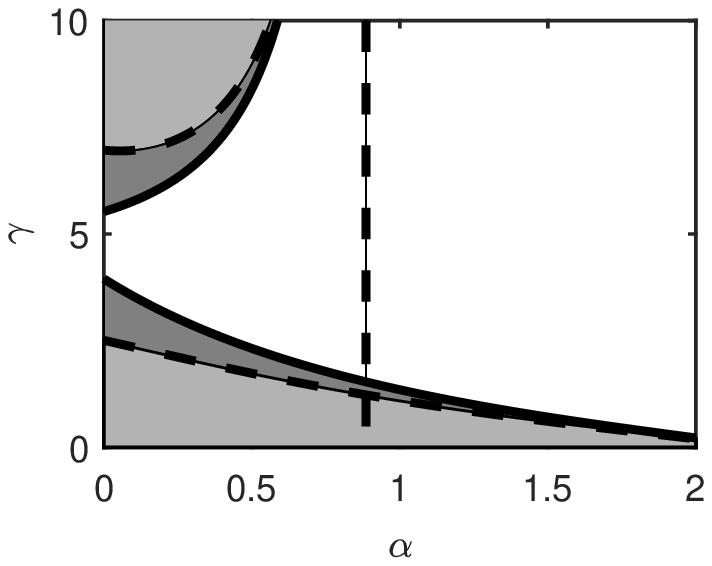}}
  \centerline{(f) }
\end{minipage}
  \caption{Friction constant $\gamma$ versus memory exponent $\alpha$ with parameters $A_0=\omega=1$, $\nu=0$, $\Omega=1.8$. (a): $\tau=0.25$, $b=0$, $\alpha_{cr1}=0.760$, $\alpha_{cr2}=1.630$; (b): $\tau=1$, $b=0$; (c): $\tau=0.55$, $b=0$, $\alpha_{cr1}=1$, $\alpha_{cr2}=1.99$; 
  (d): $\tau=0.25$, $b=0.3$, $\alpha_{cr1}=0.690$, $\alpha_{cr2}=1.720$; (e): $\tau=1$, $b=0.3$; (f): $\tau=0.55$, $b=0.3$, $\alpha_{cr1}=0.890$.  
  }\label{gama_alp_no_v}
\end{figure}

The required conditions of the appearance of critical memory exponent $\alpha_{cr}$ have been obtained from above discussions, i.e., $b<\Omega$ and $\tau\leq\tau_{max}$. Figure \ref{gama_alp_no_v} intuitively displays the resonance and non-resonance regions in the parameter space $(\gamma, \alpha)$. The dark grey domains represent the regions of the parameters $\gamma$ and $\alpha$, where SR versus $a$ is possible in the stability region, i.e., in those regions, the parameter values satisfy $0<m_{1}m-n_{1}n<a^2_{cr}$. The light grey domains are the regions where the response $A(a)$ formally also exhibits a resonance-like maximum, but in those regions the mean value $\langle x(t)\rangle$ is unstable because $a^2_{max}>a^2_{cr}$. If the critical memory exponent $\alpha_{cr}$ exists, it can be noted that with the increase of the tempering parameter $b$ $(b<\Omega)$, the critical memory exponent $\alpha_{cr1}$ (or $\alpha_{cr11}$) moves left slightly and $\alpha_{cr2}$ (or $\alpha_{cr12}$) moves right slightly; see Figure \ref{gama_alp_no_v} (a)$\rightarrow$(d), (c)$\rightarrow$(f). This conclusion can also be obtained from Figure \ref{tau_alp} (a). Next, the solid lines in Figure \ref{gama_alp_no_v} are for $\gamma_1(\alpha)$ and $\gamma_2(\alpha)$. By some calculations, we can obtain that if both $\gamma_1(0)$ and $\gamma_2(0)$ exist ($\Omega>\omega$), there are
\begin{equation}
\gamma_1(0)=\frac{\sqrt{2}(\Omega^2-\omega^2)}{(\Omega^2+b^2)^{-\frac{1}{2}}\, \Omega\,\sin(\theta-\frac{\pi}{4})}
\geq\gamma_1(0)|_{b=0}=2(\Omega^2-\omega^2)\nonumber,
\end{equation}
\begin{equation*}
\gamma_2(0)=\frac{\sqrt{2}(\Omega^2-\omega^2)}{(\Omega^2+b^2)^{-\frac{1}{2}}\, \Omega\,\sin(\theta+\frac{\pi}{4})}
\leq\gamma_2(0)|_{b=0}=2(\Omega^2-\omega^2),
\end{equation*}
and $\gamma_1(0)-\gamma_2(0)$ increases with the increase of the tempering parameter $b$. If both $\gamma_1(2)$ and $\gamma_2(2)$ exist (see Figure \ref{gama_alp_no_v}(b)(e)), there are
\begin{equation*}
\gamma_1(2)=\frac{(\omega^2-\Omega^2)\, (1+2\tau^2\, (b^2-\Omega^2)+\tau^4\, (b^2+\Omega^2)^2)}
{(\Omega^2+b^2)\, \Omega\, \tau^2\, (b-\Omega)+\Omega\, (b+\Omega)}\nonumber,
\end{equation*}
\begin{equation*}
\gamma_2(2)=\frac{(\omega^2-\Omega^2)\, (1+2\tau^2\, (b^2-\Omega^2)+\tau^4\, (b^2+\Omega^2)^2)}
{-(\Omega^2+b^2)\, \Omega\, \tau^2\, (b+\Omega)+\Omega\, (\Omega-b)}\nonumber,
\end{equation*}
\begin{equation*}
\gamma_{1,2}(2)|_{b=0}=\frac{(\omega^2-\Omega^2)\, (1-\tau^2\Omega^2)}{\Omega^2},
\end{equation*}
and $|\gamma_1(2)-\gamma_2(2)|$ increases with the increase of tempering parameter $b$. 
Besides, if both $\gamma_1(1)$ and $\gamma_2(1)$ exist ($\omega<\Omega$, $(\Omega-b)>\frac{1}{\tau}$) (see Figure \ref{gama_alp_no_v}(b)(e)), there are
\begin{equation*}
\gamma_1(1)=\frac{(\Omega^2-\omega^2)\, (1+2b\tau+\tau^2(\Omega^2+b^2))}{\Omega^2\, \tau-\Omega-\Omega\, \tau\,  b},
\end{equation*}
\begin{equation*}
\gamma_2(1)=\frac{(\Omega^2-\omega^2)\, (1+2b\tau+\tau^2(\Omega^2+b^2))}{\Omega^2\, \tau+\Omega+\Omega\, \tau\,  b},
\end{equation*}
and $\gamma_1(1)-\gamma_2(1)$ increases with the increase of the tempering parameter $b$. All in all, with the increase of the tempering parameter $b$, the resonance region in parameter space $(\gamma, \alpha)$ shrinks and the non-resonance region expands.

Next, we pay attention to a more general case, $\nu \neq0$; Figure \ref{tau_alp}(b) is related to this case. Now, the boundary lines between the regions where SR (versus  $a$) appears or not in the parameter space $(\gamma, \alpha)$ are
\begin{equation}
\gamma_{1,2}(\alpha)=\frac{2C}{-B\mp\sqrt{B^2-4AC}}
\end{equation}
with
\begin{eqnarray}
A=&f_3f_4-f_5f_6, \nonumber\\[3pt]
B=&f_1f_4(\omega^2-\Omega^2)+f_2f_3(\omega^2-\Omega^2+\nu^2)-2\Omega\nu f_2f_5, \nonumber\\[3pt]
C=&(\omega^2-\Omega^2)(\omega^2-\Omega^2+\nu^2)f_1f_2, \nonumber
\end{eqnarray}
and
\begin{eqnarray}
f_1=&(\cos(\theta\alpha)+\tau^\alpha(\Omega^2+b^2)^{\alpha/2})^2+\sin^2(\theta\alpha), \nonumber\\[3pt]
f_2=&(\cos(\varphi\alpha)+\tau^\alpha(\Omega^2+(\nu+b)^2)^{\alpha/2})^2+\sin^2(\varphi\alpha), \nonumber\\[3pt]
f_3=&(\Omega^2+b^2)^{\alpha-1}\Omega^2\tau^\alpha+(\Omega^2+b^2)^{\alpha/2-1}(\Omega^2\cos(\theta\alpha)-\Omega b\, \sin(\theta\alpha)), \nonumber\\[3pt]
f_4=&(\nu(\nu+b)+\Omega^2)(\Omega^2+(\nu+b)^2)^{\alpha-1}\tau^\alpha \nonumber\\[3pt]
  &+(\Omega^2+(\nu+b)^2)^{\alpha/2-1} ((\nu(\nu+b)+\Omega^2)\cos(\varphi\alpha)-b\Omega\, \sin(\varphi\alpha)), \nonumber\\[3pt]
f_5=&(\Omega^2+b^2)^{\alpha-1}\Omega b\tau^\alpha+(\Omega^2+b^2)^{\alpha/2-1}(\Omega b\, \cos(\theta\alpha)+\Omega^2 \sin(\theta\alpha)), \nonumber\\[3pt]
f_6=&(\Omega^2+(\nu+b)^2)^{\alpha-1}\tau^\alpha b\Omega\nonumber\\[3pt]
&+(\Omega^2+(\nu+b)^2)^{\alpha/2-1}
((\nu(\nu+b)+\Omega^2)\sin(\varphi\alpha)+b\Omega\, \cos(\varphi\alpha)).\nonumber
\end{eqnarray}
Similar to the case $\nu=0$, the expression of the critical memory exponent $\alpha_{cr}$ is
\begin{equation}
f_3f_4-f_5f_6=0.
\end{equation}
After some calculations, there exist
\begin{equation}\label{1}
\fl
\alpha_{cr1}\geq\alpha_{cr1min}= (\theta+\varphi)^{-1} \arctan\left(\frac{\Omega(\nu(\nu+b)+\Omega^2-b^2)}
{b(2\Omega^2+\nu(\nu+b))}\right)   \in(0,1),
\end{equation}
\begin{equation}\label{2}
\fl
2>\alpha_{cr2}\geq\alpha_{cr2min}=(\theta+\varphi)^{-1} \left(\arctan\left(\frac{\Omega(\nu(\nu+b)+\Omega^2-b^2)}
{b(2\Omega^2+\nu(\nu+b))}\right)+\pi \right)     >1.
\end{equation}
The analyses for (\ref{1}) and (\ref{2}) are as follows:
\begin{enumerate}
  \item $\alpha_{cr1min}\rightarrow1-\frac{\pi}{4\theta}$ and $\alpha_{cr2min}\rightarrow1+\frac{\pi}{4\theta}$ as the switching  rate $\nu\rightarrow0$, which recover the adiabatic noise case;
  \item In the very fast flippings case, $\nu\rightarrow\infty$, we have $\alpha_{cr1min}\rightarrow1$ and $\alpha_{cr2min}\rightarrow1+\frac{\pi}{\theta}>2$; that is to say, in the system with harmonic potential $V(x)=\omega^2 \frac{x^2}{2}$, there only exists $\alpha_{cr1}$ , s.t., $\gamma_1(\alpha_{cr1})\rightarrow\infty$;
  \item $\alpha_{cr1min}$ ($\alpha_{cr2min}$ ) decreases (increases) with the increase of the tempering parameter $b$; moreover, when $b=0$, there is  $\alpha_{cr1min}\rightarrow\frac{\pi}{\pi+2\textrm{arctan}(\Omega/\nu)}$, which recovers the conclusion in \cite{Laas:2011}, and $\alpha_{cr2min}\rightarrow\frac{3\pi}{\pi+2\textrm{arctan}(\Omega/\nu)}$;
  \item $\alpha_{cr1}\rightarrow\alpha_{cr1min}$ in (\ref{1}) and $\alpha_{cr2}\rightarrow\alpha_{cr2min}$ in (\ref{2}) when $\tau\rightarrow0$.
\end{enumerate}
The number of the critical memory exponent $\alpha_{cr}$ also depends on the characteristic memory time $\tau$ when $\nu\neq0$. We present this relationship in Figure \ref{tau_alp}(b), and find that
\begin{enumerate}
  \item If $0\leq\tau\leq\tau_1$, there exist $\alpha_{cr1}$ and $\alpha_{cr2}$, s.t., $\gamma_1(\alpha_{cr1})\rightarrow\infty$ and $\gamma_2(\alpha_{cr2})\rightarrow\infty$;
  \item If $\tau_2\leq\tau<\tau_{max}$, there exist $\alpha_{cr11}$ and $\alpha_{cr12}$, s.t., $\gamma_1(\alpha_{cr11})\rightarrow\infty$ and $\gamma_1(\alpha_{cr12})\rightarrow\infty$;
  \item If $\tau_1<\tau<\tau_2$, or $\tau=\tau_{max}$, there exists $\alpha_{cr1}$, s.t., $\gamma_1(\alpha_{cr1})\rightarrow\infty$;
  \item If $\tau>\tau_{max}$,
there is no $\alpha_{cr}$, s.t., $\gamma_1(\alpha_{cr})\rightarrow\infty$ or $\gamma_2(\alpha_{cr})\rightarrow\infty$,
\end{enumerate}
where
\begin{eqnarray}
\tau_1=\left(\frac{-b_1-\sqrt{b^2_1-4a_1c_1}}{2a_1}\right)^{\frac{1}{2}}, \qquad
\tau_2=\left(\frac{-b_1+\sqrt{b^2_1-4a_1c_1}}{2a_1}\right)^{\frac{1}{2}}, \nonumber\\
a_1=(\Omega^2+b^2)(\Omega^2+(\nu+b)^2)\Omega^2(\nu^2+\nu b+\Omega^2-b^2), \nonumber\\
b_1=(\Omega^2+b^2)\Omega^2(\nu^2-\nu b-\Omega^2-b^2)-(\Omega^2+(\nu+b)^2)\Omega^2(\nu^2+\nu b+\Omega^2+b^2), \nonumber\\
c_1=\Omega^2(\Omega^2-\nu^2-3b\nu-b^2). \nonumber
\end{eqnarray}
One has $\tau_2\rightarrow\left(\frac{\Omega+b}{(\Omega-b)(\Omega^2+b^2)}\right)^{\frac{1}{2}}$,  $\tau_1\rightarrow\left(\frac{\Omega-b}{(\Omega+b)(\Omega^2+b^2)}\right)^{\frac{1}{2}}$ as the switching rate $\nu\rightarrow0$, which is consistent with the adiabatic noise case. In addition, there is a trend for the lines in Figure \ref{tau_alp} to move right with the increase of switching rate $\nu$.

\begin{figure}[ht]
\begin{minipage}{0.23\linewidth}
  \centerline{\includegraphics[scale=0.5]{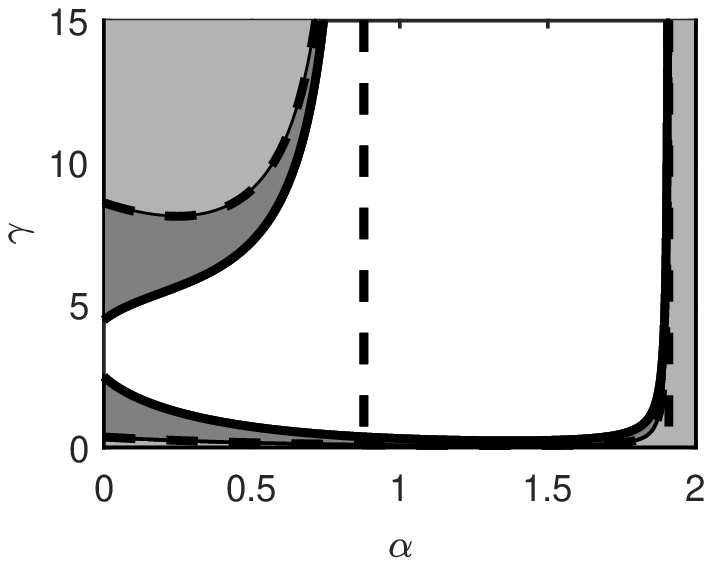}}
  \centerline{(a)}
\end{minipage}
\hfill
\begin{minipage}{0.23\linewidth}
  \centerline{\includegraphics[scale=0.5]{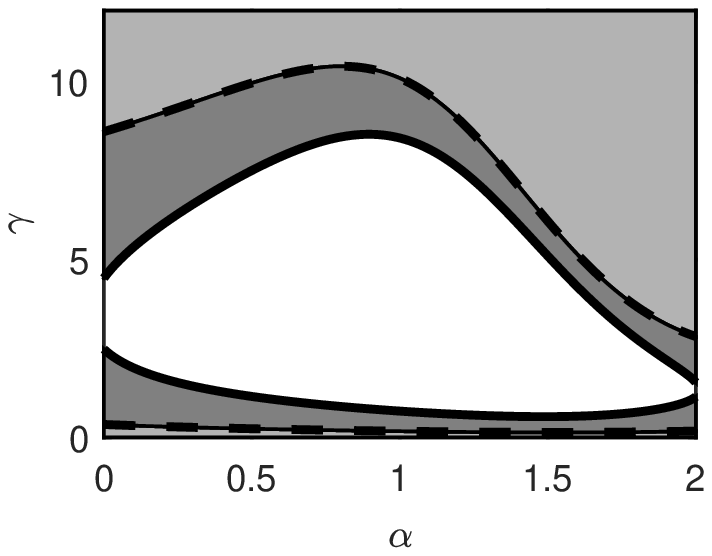}}
  \centerline{(b)}
\end{minipage}
\hfill
\begin{minipage}{0.23\linewidth}
  \centerline{\includegraphics[scale=0.5]{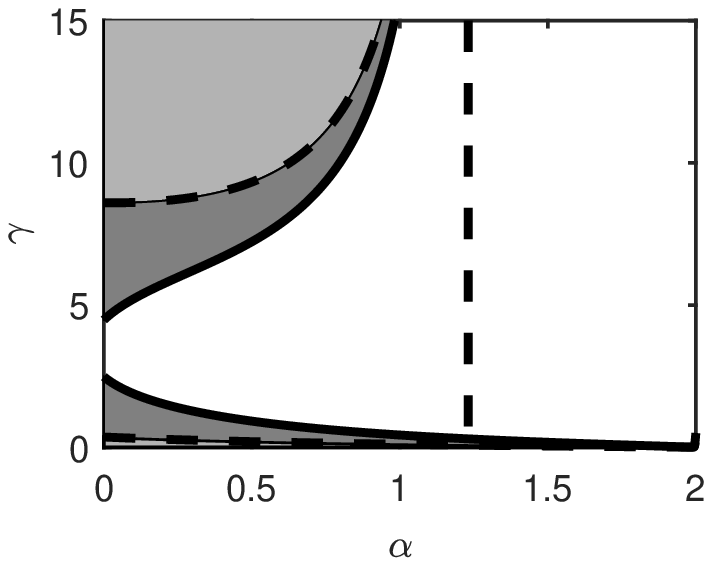}}
  \centerline{(c)}
\end{minipage}
\hfill
\begin{minipage}{0.23\linewidth}
  \centerline{\includegraphics[scale=0.5]{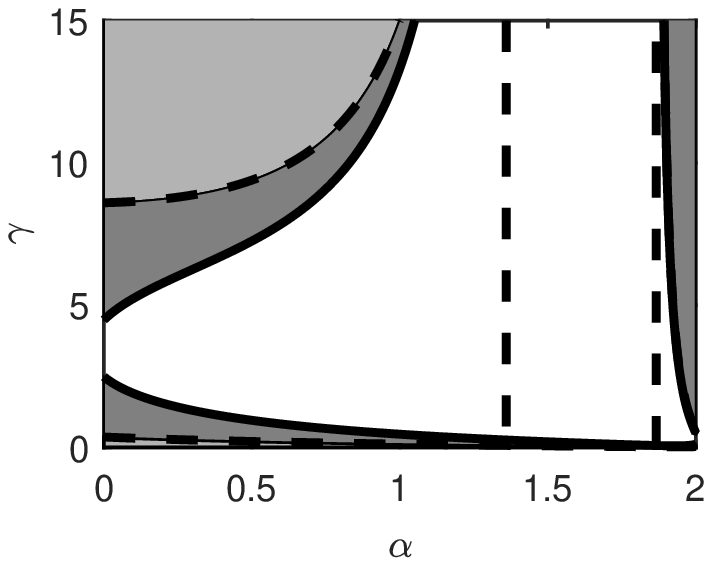}}
  \centerline{(d)}
\end{minipage}
\vfill
\begin{minipage}{0.23\linewidth}
  \centerline{\includegraphics[scale=0.5]{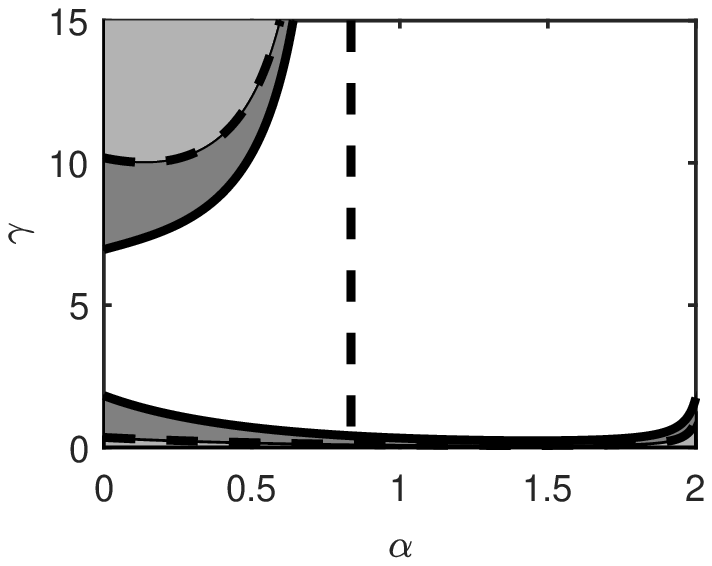}}
  \centerline{(e)}
\end{minipage}
\hfill
\begin{minipage}{0.23\linewidth}
  \centerline{\includegraphics[scale=0.5]{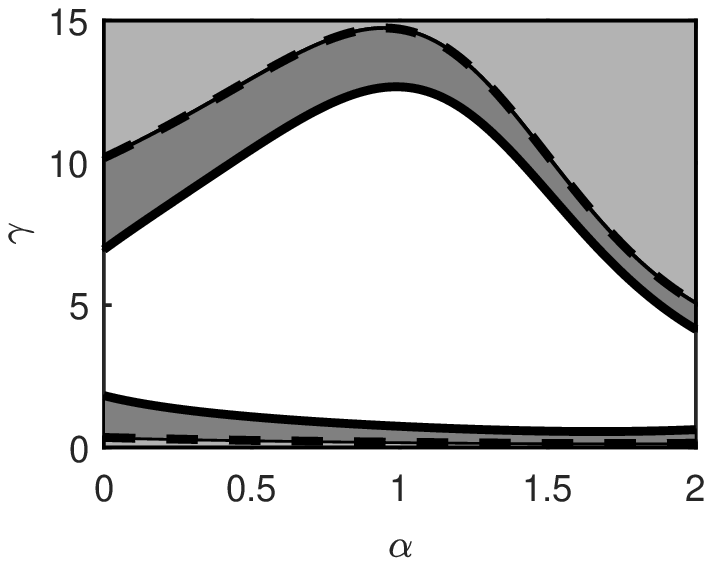}}
  \centerline{(f)}
\end{minipage}
\hfill
\begin{minipage}{0.23\linewidth}
  \centerline{\includegraphics[scale=0.5]{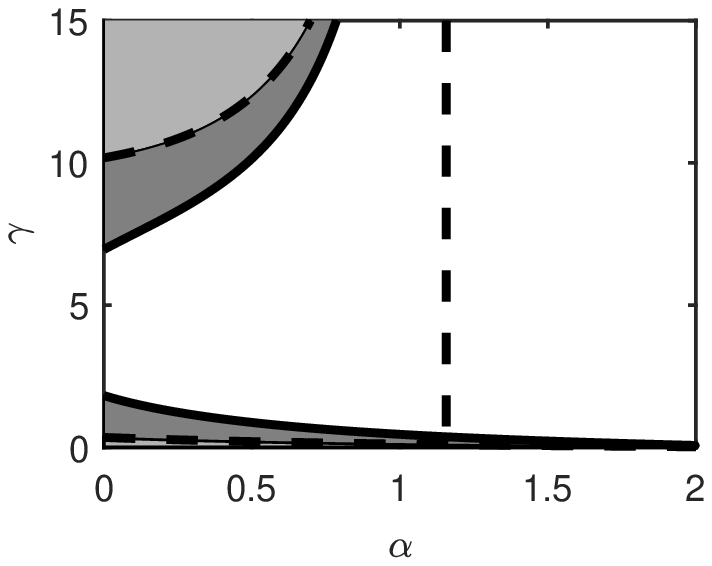}}
  \centerline{(g) }
\end{minipage}
\hfill
\begin{minipage}{0.23\linewidth}
  \centerline{\includegraphics[scale=0.5]{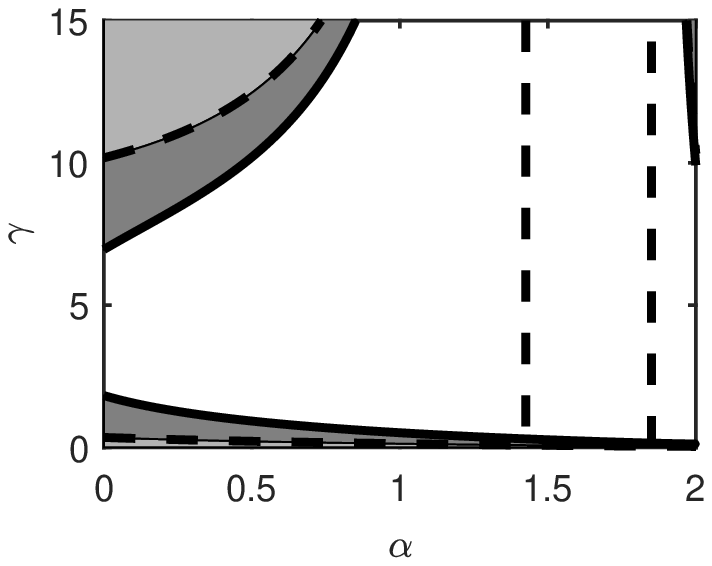}}
  \centerline{(h)}
\end{minipage}
  \caption{Friction constant $\gamma$ versus memory exponent $\alpha$ with parameters $A_0=\omega=1$, $\nu=1$, $\Omega=1.8$. (a): $\tau=0.25$, $b=0$, $\alpha_{cr1}=0.879$, $\alpha_{cr2}=1.910$; (b): $\tau=1$, $b=0$; (c): $\tau=0.55$, $b=0$, $\alpha_{cr1}=1.231$; (d): $\tau=0.6$, $b=0$, $\alpha_{cr11}=1.360$, $\alpha_{cr12}=1.867$; (e): $\tau=0.25$, $b=0.3$, $\alpha_{cr1}=0.835$; (f): $\tau=1$, $b=0.3$; (g): $\tau=0.55$, $b=0.3$, $\alpha_{cr1}=1.157$;  (h): $\tau=0.65$, $b=0.3$, $\alpha_{cr11}=1.426$, $\alpha_{cr12}=1.851$.}\label{gama_alp_have_v}
\end{figure}

In Figure \ref{gama_alp_have_v}, we depict the resonance and non-resonance regions in the parameter space $(\gamma, \alpha)$. Being the same as Figure \ref{gama_alp_no_v}, the dark grey shaded domains correspond to the regions where SR versus $a$ is possible in stability region. The light grey domains represent the unstable regions but the response $A(a)$ formally also exhibits a resonance-like maximum. With the increase of the tempering parameter $b$, the critical memory exponent $\alpha_{cr1}$ (or $\alpha_{cr11}$) moves left slightly and $\alpha_{cr2}$ (or $\alpha_{cr12}$) moves right slightly, the similar conclusion of which can also be drawn from Figure \ref{tau_alp}(b).
Similar to the case $\nu=0$, the resonance region shrinks and the non-resonance region expands with  the increase of the tempering parameter $b$. 
Moreover, when the noise switching rate $\nu$ is large and the characteristic memory time $\tau$ is small, system (\ref{SR}), without the influence of multiplicative noise, behaves as a fractional oscillatory system. In this case, there only exists one $\alpha_{cr1}\in(0, 1)$, s.t., $\gamma_1(\alpha_{cr1})\rightarrow\infty$. On the left side of $\alpha_{cr1}$, i.e., $\alpha\rightarrow0$, for large friction force $\gamma$, the resonance phenomenon still occurs.
But when $\alpha\rightarrow1$, the emergence of resonance phenomenon needs small friction force $\gamma$. This is the cage effect \cite{Burov:2008}, i.e., for small $\alpha$, the friction force is not just slowing down the particle but also causing the particle to develop a ratting motion.

\section{Signal-to-noise ratio}\label{four}
Another statistical quantity, signal-to-noise ratio (SNR), can also describe the SR phenomenon. 
In order to obtain the expression of SNR, we define $X_5=\langle x(t)x(t')\rangle$, $X_6=\langle \dot{x}(t)x(t')\rangle$, $X_7=\langle z(t)x(t)x(t')\rangle$, $X_8=\langle z(t)\dot{x}(t)x(t')\rangle$. The following expressions are obtained by the same method used in Section \ref{two}:
 \begin{eqnarray}\label{x}
    \dot{X}_5=&X_6, \nonumber\\
    \dot{X}_6=&-\omega^2X_5-X_7-\int_0^t\eta(t-u)X_6(u)\rmd u+A_0 \textrm{cos}(\Omega t)\langle x(t')\rangle, \nonumber\\
    \dot{X}_7=&-\nu X_7+ X_8, \nonumber\\
    \dot{X}_8=&-a^2X_5-\omega^2X_7-\nu X_8- \textrm{e}^{-\nu t}\int_0^t\eta(t-u)X_8(u)\textrm{e}^{\nu u}\rmd u\nonumber\\
    &+A_0 \textrm{cos}(\Omega t)\langle x(t')z(t)\rangle.
 \end{eqnarray}
To solve these equations, we use Laplace transform technique and obtain that
\begin{eqnarray}
X_5(s)=&\frac{(s+\nu)^2+(s+\nu)\eta(s+\nu)+\omega^2}{D(s)}\frac{A_0s}{s^2+\Omega^2}\langle x(t')\rangle \nonumber\\
&-\frac{A_0\mathcal{L}[\textrm{cos}(\Omega t)\langle z(t)x(t')\rangle]}{D(s)}+\sum_{i=5}^8L_i(s)X_i(0)
\end{eqnarray}
with
\begin{eqnarray*}
L_5(s)=\frac{[(s+\nu)^2+(s+\nu)\eta(s+\nu)+\omega^2][s+\eta(s)]}{D(s)},\\
L_6(s)=\frac{(s+\nu)^2+(s+\nu)\eta(s+\nu)+\omega^2}{D(s)},\\
L_7(s)=-\frac{s+\nu+\eta(s+\nu)}{D(s)},\\
L_8(s)=-\frac{1}{D(s)}.
\end{eqnarray*}
By the inverse Laplace transform, the asymptotic expression of $\langle x(t)x(t')\rangle$ in the long time limit is
\begin{eqnarray}\label{corr}
\langle x(t)x(t')\rangle_{as}=&\textrm{sgn}(\chi'_1(\Omega))\textrm{sgn}(\chi'_2(\Omega))\frac{A_1A_2}{a^2}\textrm{cos}(\Omega t+\Psi_2)\textrm{cos}(\Omega t'+\Psi_1)\textrm{e}^{-\nu|t-t'|}\nonumber\\
&+A^2\textrm{cos}(\Omega t+\Psi)\textrm{cos}(\Omega t'+\Psi),
\end{eqnarray}
where
\begin{equation*}
\chi'_2(\Omega)+\rmi \chi''_2(\Omega)=H_{32}(-\rmi \Omega-\nu),
\end{equation*}
\begin{equation*}
A_2=A_0\frac{a^2}{((m_1m_2-n_1n_2-a^2)^2+(m_1n_2+n_1m_2)^2)^{\frac{1}{2}}},
\end{equation*}
and
\begin{equation*}
\Psi_2=\textrm{arctan}\left(-\frac{m_1n_2+n_1m_2}{m_1m_2-n_1n_2-a^2}\right),
\end{equation*}
with
\begin{eqnarray*}
\varphi_2=\textrm{arctan}\left(\frac{\Omega}{b-\nu}\right),\qquad
\phi_2=\textrm{arctan}\left(-\frac{\Omega}{\nu}\right),\\ [3pt]
\chi'_2(\Omega)=-\frac{a^2(m_1m_2-n_1n_2-a^2)}{(m_1m_2-n_1n_2-a^2)^2+(m_1n_2+n_1m_2)^2},\\[3pt]
\chi''_2(\Omega)=-\frac{a^2(m_1n_2+n_1m_2)}{(m_1m_2-n_1n_2-a^2)^2+(m_1n_2+n_1m_2)^2},\\[3pt]
f_2=\frac{(\Omega^2+(b-\nu)^2)^{\frac{\alpha}{2}} \, \cos(\varphi_2-\phi_2)\, \tau^\alpha+\cos(\varphi_2(\alpha-1)+\phi_2)}
     {[\cos(\varphi_2\alpha)+\tau^\alpha\, (\Omega^2+(b-\nu)^2)^{\frac{\alpha}{2}}]^2+ \sin^2(\varphi_2\alpha)  },  \\[3pt]
g_2=\frac{[\Omega^2+(b-\nu)^2]^{\frac{\alpha}{2}} \, \sin(\phi_2-\varphi_2)\, \tau^\alpha+\sin(\varphi_2(\alpha-1)+\phi_2)}
     {[\cos(\varphi_2\alpha)+\tau^\alpha\, (\Omega^2+(b-\nu)^2)^{\frac{\alpha}{2}}]^2+ \sin^2(\varphi_2\alpha)  },  \\[3pt]
m_2=\omega^2+\nu^2-\Omega^2+(\nu^2+\Omega^2)^{\frac{1}{2}}\, (\Omega^2+(b-\nu)^2)^{\frac{\alpha-1}{2}} \, \gamma \, f_2,  \\[3pt]
n_2=-2\Omega\nu+(\nu^2+\Omega^2)^{\frac{1}{2}}\, (\Omega^2+(b-\nu)^2)^{\frac{\alpha-1}{2}} \, \gamma \, g_2.  \\
\end{eqnarray*}
The expressions of $D(s)$, $A$, $A_1$, $\Psi$, $\Psi_1$, $\chi'_1(\Omega)$, $m$, $n$, $m_1$, $n_1$ are given in Section \ref{two}. Here the stability condition $\min\limits_{0\leq s\leq\nu} D(s-\nu)>0$ is assumed to be satisfied. Define the one-time correlation function $C(\epsilon)$ as the average of two-time correlation function (\ref{corr}) over a period $T=2\pi/\Omega$ of the external driving:
\begin{eqnarray}\label{C}
C(\epsilon):&=\frac{1}{T}\int_0^T\rmd t \langle x(t)x(t+\epsilon)\rangle_{as} \nonumber\\
&=F_1\,\textrm{cos}(\Omega\epsilon)\textrm{e}^{-\nu\epsilon}+F_2\,\textrm{sin}(\Omega\epsilon)\textrm{e}^{-\nu\epsilon}
+\frac{A^2}{2}\textrm{cos}(\Omega\epsilon),
\end{eqnarray}
where
\begin{eqnarray*}
\fl F_1=\frac{A_0^2a^2}{2}\left[\frac{(m_1m-n_1n-a^2)(m_1m_2-n_1n_2-a^2)+(m_1n+n_1m)(m_1n_2+n_1m_2)}
{((m_1m-n_1n-a^2)^2+(m_1n+n_1m)^2)((m_1m_2-n_1n_2-a^2)^2+(m_1n_2+n_1m_2)^2)}\right],\nonumber\\
\fl F_2=\frac{A_0^2a^2}{2}\left[\frac{(m_1n+n_1m)(m_1m_2-n_1n_2-a^2)-(m_1m-n_1n-a^2)(m_1n_2+n_1m_2)}
{((m_1m-n_1n-a^2)^2+(m_1n+n_1m)^2)((m_1m_2-n_1n_2-a^2)^2+(m_1n_2+n_1m_2)^2)}\right].\nonumber
\end{eqnarray*}
Taking advantage of the conclusions in \cite{Gammaitoni:1998,Jung:1991,Casado-Pascual:2003}, we can rewrite (\ref{C}) as the sum of the coherent part $C_{coh}(\epsilon)$ and the incoherent part $C_{incoh}(\epsilon)$, i.e.,
\begin{equation}\label{Ctau}
C(\epsilon)=C_{coh}(\epsilon)+C_{incoh}(\epsilon),
\end{equation}
where
\begin{eqnarray}\label{Ccoh}
C_{coh}(\epsilon):=\frac{1}{T}\int_0^T \rmd t \langle x(t)\rangle_{as}\langle x(t+\epsilon)\rangle_{as}
=\frac{A^2}{2}\textrm{cos}(\Omega\epsilon)
\end{eqnarray}
and
\begin{eqnarray*}
C_{incoh}(\epsilon)&:=\frac{1}{T}\int_0^T\rmd t (\langle x(t)x(t+\epsilon)\rangle_{as}-\langle x(t)\rangle_{as}\langle x(t+\epsilon)\rangle_{as})\\
&~=F_1\,\textrm{cos}(\Omega\epsilon)\textrm{e}^{-\nu\epsilon}+F_2\,\textrm{sin}(\Omega\epsilon)\textrm{e}^{-\nu\epsilon}.
\end{eqnarray*}
The output SNR is defined as \cite{Gammaitoni:1998,Mankin:2008,Casado-Pascual:2003,McNamara:1989}
\begin{equation}\label{SNR}
\textrm{SNR}:=\frac{\lim\limits_{\varepsilon \to 0^+}\int_{\Omega-\varepsilon}^{\Omega+\varepsilon}\rmd k\, \tilde{C}(k)}{\tilde{C}_{incoh}(\Omega)}=\frac{\Gamma_1}{\Gamma_2},
\end{equation}
where $\tilde{C}(k):= \frac{1}{\pi}\int_{-\infty}^{+\infty}\textrm{e}^{-\rmi k\epsilon}C(\epsilon)\rmd\epsilon$ is the Fourier transform of $C(\epsilon)$. Using Fourier cosine transform, the expressions of $\Gamma_1$ and $\Gamma_2$ are as follows
\begin{eqnarray}\label{gama}
\Gamma_1&=\frac{2}{T}\int_0^T\rmd\epsilon\, C_{coh}(\epsilon) \,\textrm{cos}(\Omega\, \epsilon)=\frac{A^2}{2},\\
\Gamma_2&=\frac{2}{\pi}\int_0^ \infty \rmd\epsilon\, C_{incoh}(\epsilon)\, \textrm{cos}(\Omega\, \epsilon)=\frac{2}{\pi}
\left(\frac{F_1(\nu^2+2\Omega^2)+F_2\nu\Omega}{\nu(\nu^2+4\Omega^2)}\right).
\end{eqnarray}
Finally, the output SNR at the driving frequency $\Omega$ is
\begin{equation}
\textrm{SNR}
=\frac{A^2\pi\nu(\nu^2+4\Omega^2)}{4(F_1(\nu^2+2\Omega^2)+F_2\, \nu\, \Omega)}.
\end{equation}

\begin{figure}[ht]
\begin{minipage}{0.52\linewidth}
  \centerline{\includegraphics[scale=0.3]{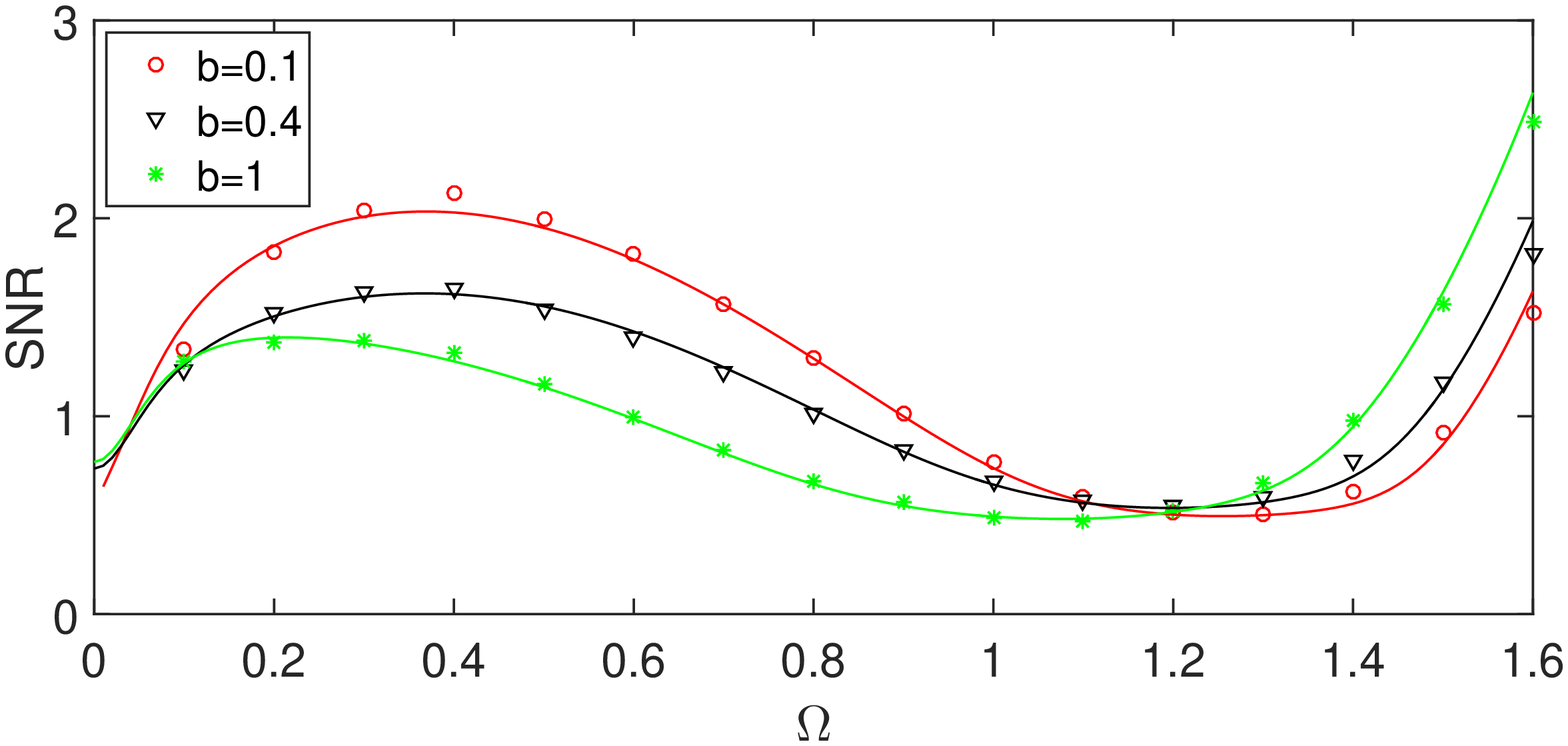}}
  \centerline{(a)}
\end{minipage}
\hfill
\begin{minipage}{0.49\linewidth}
  \centerline{\includegraphics[scale=0.3]{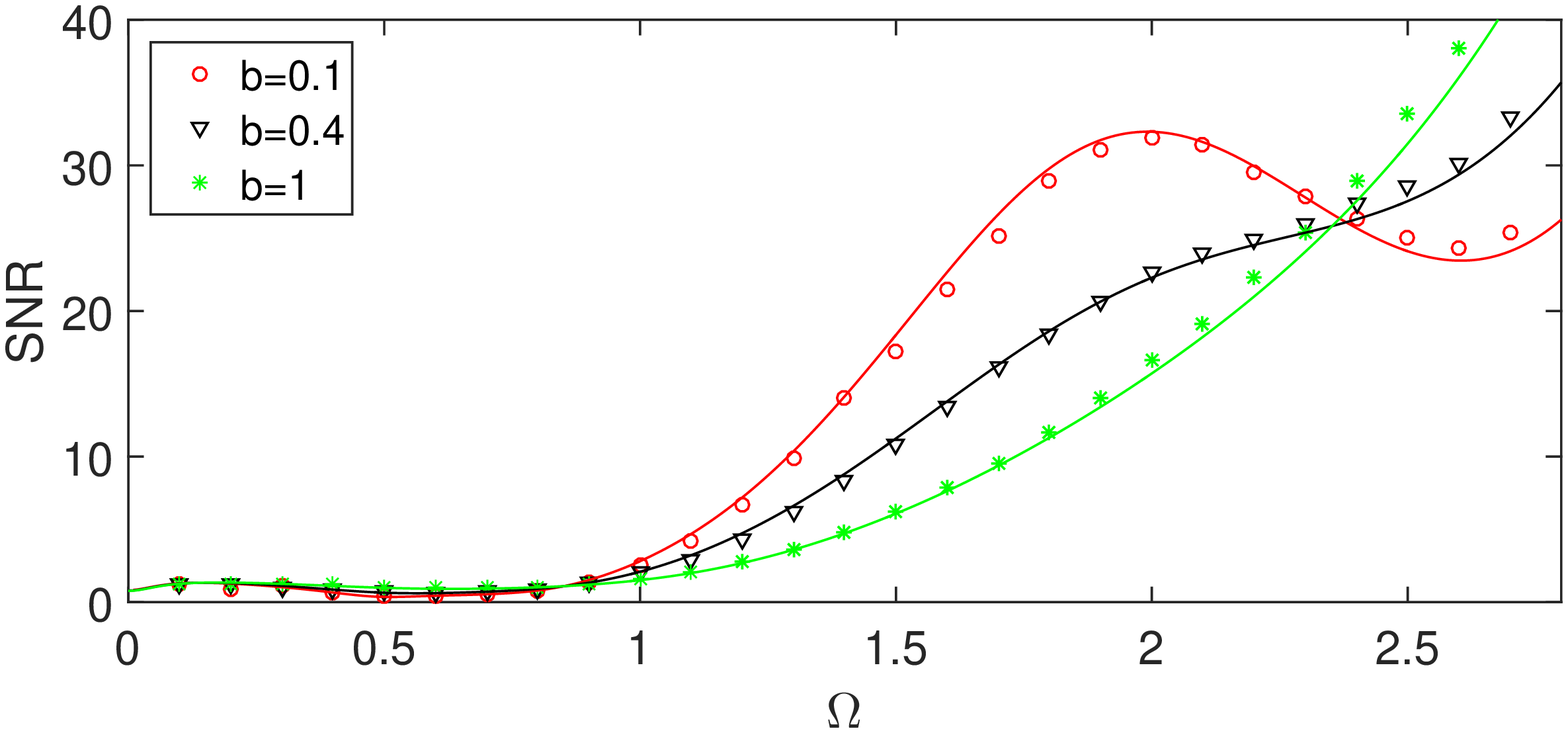}}
  \centerline{(b)}
\end{minipage}
\caption{SNR versus driving frequency $\Omega$ with parameters $A_0=\omega=1$, $\tau=0.5$, $\gamma=1$, $\nu=0.1$, $a^2=0.2$. (a): $\alpha=0.7$; (b): $\alpha=1.6$. Solid lines are the analytical results, and the marks are the computer simulations with the time $T = 150$ and the number of sampled trajectories 7000.}\label{SNR_Omg}
\end{figure}

SNR versus driving angular frequency $\Omega$, for different tempering parameter $b$ and memory exponent $\alpha$, is shown in Figure \ref{SNR_Omg}, the non-monotonic dependence of which is so-called bona fide resonance. In Figure \ref{SNR_Omg}(a) ($\alpha=0.7$), with the increase of the tempering parameter $b$, the resonance peak and the resonance valley move left slightly. The value of resonance peak decreases, while the value of resonance valley is almost unaffected by the tempering parameter $b$. In Figure \ref{SNR_Omg}(b) ($\alpha=1.6$), with the increase of tempering  parameter $b$, the resonance peak suffers a suppression, while the resonance valley rises up. Finally, there is a monotonic behavior of SNR versus driving frequency $\Omega$.

\section{Conclusion}\label{five}
Stochastic resonance has a big potential in applications, e.g., detecting signal with information from noisy enviroment. This paper discusses the SR in the generalized Langevin system with the tempered Mittag-Leffler memory kernel. Besides internal noise, the influence of the fluctuating environment is modeled by the multiplicative dichotomous noise. Using the Shapiro-Loginov formula, we get the exact expressions of the first moment and the correlation function of the output signal as well as SNR of the stochastic oscillator system.
   The obtained results for this oscillator dynamical system with the tempered Mittag-Leffler memory kernel for the friction term could recover the ones of the system with Mittag-Leffler memory kernel, power-law memory kernel, and exponent-form memory kernel. We analyze the SR phenomena, being the non-monotonic behavior of the output amplitude and SNR, and find
that the exponential tempering exerts an influence on the critical memory exponent and shrinks the resonance region in parameter space $(\gamma, \alpha)$. The extensively performed numerical simulations verify the theoretical results.


\section*{Acknowledgments}
This work was supported by the National Natural Science Foundation of China under Grant No. 11671182, and the Fundamental Research Funds for the Central Universities under Grant No. lzujbky-2017-ot10.

\appendix
\setcounter{section}{0}
\section{ Algorithms for numerical simulations}

There are many numerical simulations provided above, which boil down to two parts, i.e., the asymptotic expression of the first moment of position $\langle x(t)\rangle_{as}$ and SNR. For the first part, it is sufficient to generate the trajectories of $x(t)$ in (\ref{SR}). To do this, we must deal with the noise $z(t)$ and $\xi(t)$ first. Note that the additive noise $\xi(t)$ is independent from $z(t)$ and has a zero mean. In this work, since we are only concerned about the behavior of the first moment $\langle x(t)\rangle_{as}$, the term $\xi(t)$ can be omitted in numerical simulations for reducing the computation cost. As to the dichotomous noise $z(t)$, it can only take the values between $a$ and $-a$, and is completely characterized by the transition probability
\begin{eqnarray*}
P_{ij}(\tau)&=
\frac{1}{2}\left(\begin{array}{cc} 1+\textrm{e}^{-\nu \tau}  &  1-\textrm{e}^{-\nu \tau} \\ 1-\textrm{e}^{-\nu \tau}  & 1+\textrm{e}^{-\nu \tau} \end{array} \right), ~~~i, j\in\{-a, a\},
\end{eqnarray*}
where $\tau$ is the time step.
Assume that the initial distribution of $z(t)$ is uniform, i.e., $P(z(t_0)=a)=P(z(t_0)=-a)=\frac{1}{2}$. From the transition probability, we find that $z(t)$ has the probability $(1+\textrm{e}^{-\nu \tau})/2$ to keep its value and the probability $(1-\textrm{e}^{-\nu \tau})/2$ to change its value at each step. Therefore, at the $j$-th step, we can generate a random number $r_j$ with uniform distribution in $[0,1]$, and then $z(t_{j+1})=z(t_j)\cdot\textrm{sgn}(r_j-(1-\textrm{e}^{-\nu \tau})/2)$. Step by step, the dichotomous noises $z(t_0),z(t_1),\cdots$, are obtained.

Then we solve (\ref{SR}) with the scheme
\begin{equation*}
\fl
\label{cases}
\cases{x(t_{j+1})=x(t_j)+v(t_j)\cdot\tau, \\
\frac{v(t_{j+1})-v(t_j)}{\tau}=
    -\int_0^{t_{j+1}}\eta(t_{j+1}-t')\,v(t')\rmd t'-w^2\,x(t_{j+1})-z(t_j)\,x(t_{j+1})+A_0\cos(\Omega\,t_j),\\}
\end{equation*}
where
\begin{equation}\label{Trapform}
\fl
\int_0^{t_{j+1}}\eta(t_{j+1}-t')\,v(t')\rmd t' =\frac{\tau}{2}(\eta(t_{j+1})\,v(t_0)+\eta(t_0)\,v(t_{j+1}))+\tau\sum_{k=1}^{j}\eta(t_k)\,v(t_{j+1-k}).
\end{equation}
By this scheme, we can generater $N$ trajectories of $x^{(i)}(t_n),i=1,\cdots,N$, and then obtain
\begin{equation*}
  \langle x(t_n)\rangle \approx \frac{1}{N}\sum_{i=1}^N x^{(i)}(t_n).
\end{equation*}

Next, we use $x^{(i)}(t_n)$ obtained above to compute SNR. We can calculate formula (\ref{Ctau}) and (\ref{Ccoh}), i.e.,

\begin{equation*}
\label{cases}
    \cases{C(\epsilon)= \frac{1}{T}\int_0^T \langle x(t+\epsilon)x(t)\rangle_{as} \,\rmd t,  \\
    C_{coh}(\epsilon)= \frac{1}{T}\int_0^T \langle x(t+\epsilon)\rangle_{as} \langle x(t)\rangle_{as} \,\rmd t,  \\
    C_{incoh}(\epsilon)= C(\epsilon)-C_{coh}(\epsilon),\\}
\end{equation*}
by
\begin{eqnarray*}
    &\langle x(t_n+\epsilon)x(t_n)\rangle \approx \frac{1}{N}\sum_{i=1}^N x^{(i)}(t_n+\epsilon)\cdot x^{(i)}(t_n), \\
    &\langle x(t_n+\epsilon)\rangle \langle x(t_n)\rangle \approx
    \left(\frac{1}{N}\sum_{i=1}^N x^{(i)}(t_n+\epsilon)\right)\cdot \left(\frac{1}{N}\sum_{i=1}^N x^{(i)}(t_n)\right).
\end{eqnarray*}
The notation $\langle\cdot\rangle_{as}$ denotes the asymptotic behaviour of $\langle x(t)\rangle$, so we only use the later parts of $x^{(i)}(t_n)$ to obtain the corresponding results. Then we use trapezoidal formula to compute $C(\epsilon),C_{coh}(\epsilon)$ like (\ref{Trapform}), where for convenience we take $\epsilon$ to be an integer multiple of time step $\tau$.
Following this, SNR can be obtained by the formulae (\ref{SNR}) and (\ref{gama}), i.e.,
\begin{equation*}
\label{cases}
    \cases{\Gamma_1= \frac{2}{T}\int_0^T C_{coh}(\epsilon)\,\cos(\Omega\,\epsilon)\,\rmd\epsilon \\
    \Gamma_2= \frac{2}{\pi}\int_0^\infty C_{incoh}(\epsilon)\,\cos(\Omega\,\epsilon)\,\rmd\epsilon \\
    \textrm{SNR}=\Gamma_1/\Gamma_2.}
\end{equation*}
Since $C_{incoh}(\epsilon)$ decays to $0$ for large values of $\epsilon$, the integral interval of $\Gamma_2$ need to be truncated first. Then the integrals of $\Gamma_1$ and $\Gamma_2$ can be  numerically computed by trapezoidal formula.

\end{document}